\documentclass[onecolumn,letterpaper,11pt,accepted=2022-12-19]{quantumarticle}

\pdfoutput=1
\usepackage[numbers,sort&compress,square]{natbib}

\usepackage{amssymb,amsmath,amsthm,amsfonts,amscd}
\usepackage{graphicx,xcolor}
\usepackage{hyperref} 
\usepackage[ruled]{algorithm2e}
\usepackage{extarrows}

\DeclareMathOperator*{\argmin}{arg\,min}
\newcommand{\face}[2]{{\Delta_{#1}(#2)}}
\newcommand{\facex}[3]{{\Delta^{#1}_{#2}(#3)}}
\newcommand{\link}[2]{{\mathrm{Lk}_{#1}(#2)}}
\newcommand{\linkx}[3]{{\mathrm{Lk}^{#1}_{#2}(#3)}}
\renewcommand{\star}[2]{{\mathrm{St}_{#1}(#2)}}
\DeclareMathOperator{\im}{im}
\newcommand{\bnd}[2]{\partial_{#1,#2}}
\newcommand{\rest}[1]{|_{#1}}
\newcommand{\col}[1]{{\mathrm{color}(#1)}}

\newcommand{\psqoct}{10.2\%}
\newcommand{\psqoctUF}{9.8\%}

\renewcommand{\phi}{\varphi}

\newtheorem{definition}{Definition}
\theoremstyle{definition}

\newtheorem{lemma}{Lemma}
\theoremstyle{lemma}

\newtheorem{theorem}{Theorem}
\theoremstyle{theorem}

\theoremstyle{corollary}

\begin{document}

\title{Efficient color code decoders in $d\geq 2$ dimensions from toric code decoders}

\author{Aleksander Kubica}
\affiliation{Perimeter Institute for Theoretical Physics, Waterloo, ON N2L 2Y5, Canada}
\affiliation{Institute for Quantum Computing, University of Waterloo, Waterloo, ON N2L 3G1, Canada}
\author{Nicolas Delfosse}
\affiliation{Station Q Quantum Architectures and Computation Group, Microsoft Research, Redmond, WA 98052, USA}

\begin{abstract}
We introduce an efficient decoder of the color code in $d\geq 2$ dimensions, the Restriction Decoder, which uses any $d$-dimensional toric code decoder combined with a local lifting procedure to find a recovery operation.
We prove that the Restriction Decoder successfully corrects errors in the color code if and only if the corresponding toric code decoding succeeds.
We also numerically estimate the Restriction Decoder threshold for the color code in two and three dimensions against the bit-flip and phase-flip noise with perfect syndrome extraction.
We report that the 2D color code threshold $p_{\textrm{2D}} \approx \psqoct$ on the square-octagon lattice is on a par with the toric code threshold on the square lattice.
\end{abstract}

\maketitle


One of the leading approaches to building scalable quantum computers is based on the two-dimensional toric code \cite{Kitaev2003, Bravyi1998, Dennis2002} and exhibits the following desirable features: only geometrically local syndrome measurements are required, the accuracy threshold is high, and the overhead of fault-tolerant implementation of logical Clifford gates is low.
However, due to geometric locality of stabilizer generators, there are certain limitations \cite{Zeng2011,Eastin2009,Bravyi2013, Pastawski2014,Jochym-OConnor2018} on ``easy-to-implement'' logical gates.
In particular, it is not known how to fault-tolerantly implement a non-Clifford gate with low overhead in the 2D toric code.
The formidable qubit overhead of state distillation \cite{Bravyi2005, Fowler2012} associated with the necessary non-Clifford gate motivates the search for alternative ways of achieving universality.

The color code~\cite{Bombin2006,Bombin2007,Kubicathesis} provides an alternative to the toric code that may make fault-tolerant computation on encoded qubits easier.
Namely, in the 2D color code one can transversally implement the logical Clifford group, and by code switching and gauge fixing \cite{Paetznick2013, Anderson2014} between the 2D and 3D versions of the color code one can fault-tolerantly implement the logical non-Clifford $T = \textrm{diag}(1,e^{i \pi/4})$ gate~\cite{Bombin2016, Bombin2013, Kubica2015a,Beverland2021}.
Unfortunately, despite a close connection between the color and toric codes~\cite{Kubica2015}, the color code is generally considered to be more difficult to decode  than the toric code~\cite{Wang2010,Landahl2011}.
In particular, the highest reported error-correction thresholds for the 2D color code with efficient decoding algorithms are around $7.8\%\sim 8.7\%$~\cite{Sarvepalli2012,Bombin2012a, Delfosse2014}.
Those threshold values are clearly below the thresholds of $9.9\% \sim 10.3\%$ for leading efficient toric code decoders, which include the Minimum-Weight Perfect Matching (MWPM) algorithm~\cite{Dennis2002} and the Union-Find (UF) decoder~\cite{Delfosse2017}.
This seemingly suboptimal performance of the color code is one of the reasons why the 2D toric code, not the 2D color code, is the leading quantum computation approach.

In our work we show that the color code performance is actually on a par with the toric code.
First we consider the two-dimensional case which is the most relevant for future experiments.
We propose the Restriction Decoder for the 2D color code, which uses toric code decoding with a local lifting procedure to find a recovery operator.
The Restriction Decoder combines and advances the ideas from Refs.~\cite{Bombin2012a,Delfosse2014} about how to decode the 2D color code.
We also estimate the Restriction Decoder threshold for the 2D color code against the bit-flip and phase-flip noise with perfect syndrome extraction to be $p_{\textrm{2D}} \approx \psqoct$.
Importantly, this threshold matches the MWPM algorithm threshold for the toric code, as well as the highest previously reported 2D color code thresholds achieved by heuristic neural-network~\cite{Maskara2018} and tensor-network decoding~\cite{Tuckett2019}. 
Our result therefore indicates that the quantum computation approach based on the 2D color code constitutes a competitor to the one based on the 2D toric code, however further detailed investigation of circuit-level thresholds would be very desired.

We then proceed with generalizing the Restriction Decoder to be applicable to the color code in $d\geq 2$ dimensions.
Similarly as in two dimensions, we build on the idea that the problem of color code decoding can be reduced to the problem of toric code decoding, which has been extensively studied~\cite{Dennis2002,Pastawski2011,Duivenvoorden2017,Breuckmann2017,Breuckmann2017a,Kubica2018toom}
In our construction we rely on:
(i) a morphism of chain complexes of the $d$-dimensional color and toric codes, which maps the structure of one code onto the other,
(ii) a local lifting procedure that estimates a color code recovery operator from the toric code correction.
The Restriction Decoder uses any $d$-dimensional toric code decoder and a local lifting procedure, and thus can constitute an efficient solution to the problem of decoding the color code in $d\geq 2$ dimensions.
In particular, the time complexity of the Restriction Decoder is proportional to the time complexity of the toric code decoder that we use.
We prove that the Restriction Decoder successfully corrects errors in the color code if and only if the corresponding toric code decoding succeeds.
This in turn allows us to estimate the Restriction Decoder threshold from the toric code threshold.

The structure of the article is as follows.
We start by explaining the Restriction Decoder for the 2D color code in Sec.~\ref{sec_2d}.
Then, in Sec.~\ref{sec_basics} we mostly review basic ideas, including CSS stabilizer codes, chain complexes, lattices and the problem of toric code decoding.
In Sec.~\ref{sec_lattices} we discuss the notions of colexes and restricted lattices, as well as prove some technical lemmas, which are later needed in Sec.~\ref{sec_decoding} to generalize the Restriction Decoder to $d\geq 2$ dimensions.
Since those two sections contain all the technical details and proofs, some readers may choose to skip them and proceed directly to Sec.~\ref{sec_examples}, where we illustrate the Restriction Decoder with more examples in two and three dimension and provide numerical estimates of the threshold.

\section{High-threshold decoder for the 2D color code}
\label{sec_2d}

In this section we discuss the two-dimensional color code and the problem of color code decoding.
Then, we describe the Restriction Decoder for the 2D color code as well as provide the numerical estimates of its threshold.
For technical details about the Restriction Decoder we refer the reader to Sec.~\ref{sec_restriction_decoder}, where we analyze its general $d$-dimensional version for $d\geq 2$.

\subsection{A brief overview of the 2D color code}

Stabilizer codes are quantum error correcting codes \cite{Shor1995, Gottesman1996} specified by the stabilizer group $\mathcal{S}$, which is an Abelian subgroup of the Pauli group $\mathcal{P}_n$ generated by tensor products of Pauli operators on $n$ qubits.
The code space associated with the stabilizer group $\mathcal{S}$ is spanned by $+1$ eigenvectors of stabilizers $S\in\mathcal{S}$, and thus for the code space to be non-trivial we require $-I\not\in\mathcal{S}$.
We focus our emphasis on CSS stabilizer codes \cite{Calderbank1996,Calderbank1997}, whose stabilizer group is generated by $X$- and $Z$-type stabilizer generators.
Since for any CSS stabilizer code one can independently correct Pauli $X$ and $Z$ errors, in what follows we concentrate on correcting $Z$ errors; correcting $X$ errors can be done analogously.

The 2D color code is an example of a topological CSS stabilizer code~\cite{Bombin2013book}, which is defined on a two-dimensional lattice $\mathcal{L}$ built of triangles; see Fig.~\ref{fig_colorcode_2D}(a).
Importantly, we require that the vertices of $\mathcal{L}$ are $3$-colorable, i.e., we can assign three colors $R$, $G$ and $B$ to the vertices in such a way that any two vertices incident to the same edge have different colors.
Let us denote by $\face 0 {\mathcal{L}}$, $\face 1 {\mathcal{L}}$ and $\face 2 {\mathcal{L}}$ the sets of vertices, edges and triangular faces of the lattice $\mathcal{L}$, respectively.
We place one qubit on every face $f\in\face 2 {\mathcal{L}}$.
The $X$- and $Z$-type stabilizer generators $S_X(v)$ and $S_Z(v)$ of the 2D color code are associated with every vertex $v\in\face 0 {\mathcal{L}}$.
They are defined as products of Pauli $X$ and $Z$ operators on qubits on all the faces $f\in\face 2 {\mathcal{L}}$ neighboring 
the vertex $v$, i.e., $S_X(v) = \prod_{f \ni v} X(f)$ and $S_Z(v) = \prod_{f \ni v} Z(f)$. 
Logical Pauli $X$, $Y$ and $Z$ operators for the 2D color code can be implemented by string-like $X$, $Y$ and $Z$ operators forming non-contractible loops.

\subsection{The problem of color code decoding}

\begin{figure}[ht!]
\centering
(a)\includegraphics[height = 0.222 \textheight]{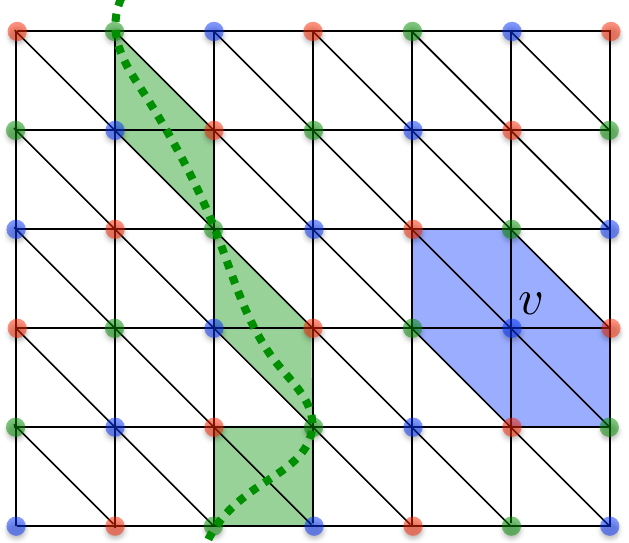}
\hspace*{12mm}
(b)\includegraphics[height = 0.22 \textheight]{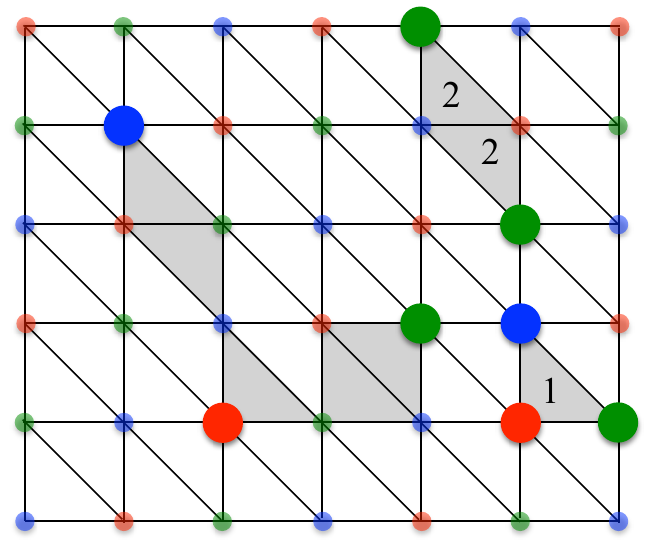}\\
(c)\includegraphics[height = 0.22 \textheight]{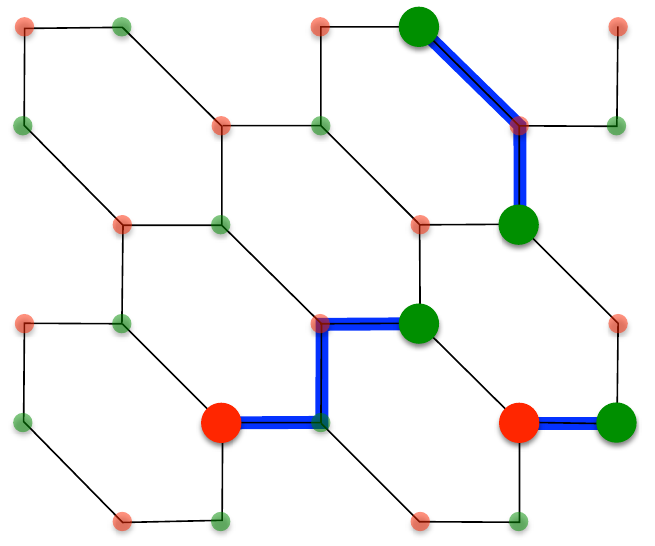}
\hspace*{10mm}
(d)\includegraphics[height = 0.227 \textheight]{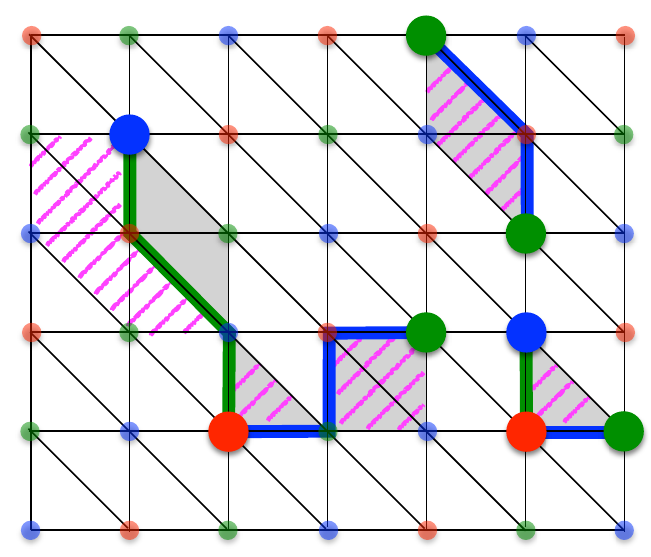}\\
\caption{
(a) Qubits of the 2D color code are placed on the triangular faces of a two-dimensional lattice $\mathcal{L}$ with $3$-colorable vertices, whereas $X$- and $Z$-type stabilizer generators $S_X(v)$ and $S_Z(v)$ (shaded in blue) are associated with every vertex $v\in \face 0 {\mathcal{L}}$.
A logical Pauli $X$, $Y$ or $Z$ operator can be implemented by a string-like operator supported on faces shaded in green.
(b) Qubits on a subset of faces $\epsilon\subseteq\face 2 {\mathcal{L}}$ (shaded in grey) are affected by Pauli $Z$ errors, which result in the $X$-type syndrome $\sigma\subseteq\face 0 {\mathcal{L}}$ (highlighted vertices).
Observe that a single-qubit error (marked by 1) creates a triple of excitations of three different colors, whereas a two-qubit error (marked by 2) creates a pair of excitations of the same color.
(c) The restricted lattice $\mathcal{L}_{RG}$ is obtained from $\mathcal{L}$ by removing all $B$ vertices of $\mathcal{L}$, as well as all the edges and faces incident to the removed vertices.
One can always find a subset of edges $\rho_{RG}\subseteq\face 1 {\mathcal{L}_{RG}}$ (blue), whose $0$-boundary matches the restricted syndrome $\sigma_{RG}\subseteq\face 0 {\mathcal{L}_{RG}}$, i.e., $\partial_1 \rho_{RG} = \sigma_{RG}$.
(d) The Restriction Decoder finds a color code correction $\phi \subseteq \face 2 {\mathcal{L}}$ (hatched in magenta) from $\rho_{RG}\cup\rho_{RB}$ (thick blue and green edges) by using a local lifting procedure.
Namely, for every $R$ vertex $v\in\facex R 0 {\rho_{RG}\cup\rho_{RB}}$ one finds a subset of neighboring faces $\tau_v\subseteq\star 2 v$, whose $1$-boundary locally matches $\rho_{RG}\cup\rho_{RB}$, i.e.,
$(\partial_2\tau_v)\rest v = (\rho_{RG}\cup\rho_{RB})\rest v$.
Note that the initial error $\epsilon$ and the correction $\phi = \bigcup_{v\in\facex R 0 {\rho_{RG}\cup\rho_{RB}}} \tau_v$ are not the same; rather, they form a stabilizer.
}
\label{fig_colorcode_2D}
\end{figure}

Let $\epsilon\subseteq \face 2 {\mathcal{L}}$ be the set of faces identified with qubits affected by $Z$ errors and $\sigma\subseteq \face 0 {\mathcal{L}}$ be the corresponding $X$-type syndrome, i.e., the set of all the $X$-type stabilizer generators anticommuting with the error $\epsilon$.
We refer to the violated stabilizer generators as 0D point-like excitations; see e.g. \cite{Maskara2018} for a comprehensive discussion.
In the 2D color code there are three species of excitations labelled by one of three colors $R$, $G$ and $B$.
Observe that a two-qubit Pauli $Z$ error on a pair of adjacent qubits leads to a pair of excitations of the same color, whereas a single-qubit Pauli $Z$ error creates a triple of excitations of all three colors; see Fig.~\ref{fig_colorcode_2D}(b).
Note that the presence of any excitations in the system indicates that the encoded logical information suffered from some errors.

In order to remove all the excitations and subsequently return the encoded information back to the code space we need an efficient decoder.
A decoder is a classical algorithm, which as the input takes a syndrome $\sigma$, i.e., an observed configuration of the excitations, and as the output returns an appropriate Pauli correction.
Since the correction is a $Z$-type operator (as we want to correct $Z$ errors), thus to describe that operator we just need to specify its support $\phi\subseteq\face 2 {\mathcal{L}}$.
We say that decoding succeeds if the initial error $\epsilon$ combined with the correction $\phi$ implements a trivial logical operator.
For concreteness, in the rest of the paper we assume the phase-flip noise, which independently affects each qubit with Pauli $Z$ error with probability $p$.
Let us consider some code family defined on lattices labeled by their linear size $L$.
We say that a decoder for the code (family) has a threshold $p_{\textrm{th}}$ if the probability of unsuccessful decoding $p_{\textrm{fail}}(p,L)$ goes to zero in the limit $L\rightarrow \infty$ as long as $p < p_{\textrm{th}}$.

One possible decoding strategy for the 2D color code is to find the most likely error consistent with the observed syndrome.
This task, however, goes beyond solving the minimum-weight perfect matching (MWPM) problem\footnote{
The Minimum-Weight Perfect Matching problem is defined as follows: \emph{for a weighted graph $G=(V,E)$ find a subset of edges $E'\subseteq E$ of minimum total weight, such that each vertex in $V$ is incident to exactly one edge in $E'$}.}
as the color code excitations are not only created or removed in pairs but also in triples.
Also, we emphasize that no decoder can simply ignore colors of the excitations since one can only  remove either pairs of excitations of the same color or triples of excitations with three different colors.
The aforementioned difficulties led to a belief that decoding of the 2D color code is more challenging than decoding of the 2D toric code \cite{Wang2010,Landahl2011}.

\subsection{The Restriction Decoder for the 2D color code}
\label{sec_decoder_2D}

Before we describe the main idea behind the Restriction Decoder for the 2D color code, we need to discuss a few concepts, such as boundaries and restricted lattices.
Let $\alpha\subseteq\face 2 {\mathcal{L}}$ be a subset of faces of the lattice $\mathcal{L}$.
We denote by $\partial_2 \alpha \subseteq \face 1 {\mathcal{L}}$ the set of all the edges of $\mathcal{L}$, which belong to an odd number of faces of $\alpha$.
We refer to $\partial_2 \alpha$ as the $1$-boundary of $\alpha$.
Similarly, if $\beta\subseteq \face 1 {\mathcal{L}}$ is a subset of edges of $\mathcal{L}$, then we define its $0$-boundary $\partial_1 \beta\subseteq \face 0 {\mathcal{L}}$ to be the set of all the vertices of $\mathcal{L}$, which are incident to an odd number of edges of $\beta$.
Finally, the restricted lattice $\mathcal{L}_{RG}$ is obtained from the lattice $\mathcal{L}$ by removing all the vertices of color $B$ as well as all the edges and faces incident to the removed vertices; see Fig.~\ref{fig_colorcode_2D}(c).
In other words, the restricted lattice $\mathcal{L}_{RG}$ only contains vertices of color $R$ or $G$, as well as edges between them; the restricted lattice $\mathcal{L}_{RB}$ is defined similarly.

Recall that the goal of color code decoding is to find a correction $\phi \subseteq \face 2 {\mathcal{L}}$, which is a collection of two-dimensional faces, from a syndrome $\sigma \subseteq \face 0 {\mathcal{L}}$, which is a configuration of zero-dimensional excitations.
By definition, $\sigma$ is the set of all the vertices of $\mathcal{L}$, which belong to an odd number of faces of the error $\epsilon\subseteq\face 2 {\mathcal{L}}$.
Let $\sigma_{RG}\subseteq\face 0 {\mathcal{L}_{RG}}$ denote the set of all the $R$ and $G$ excitations of $\sigma$.
We can view $\sigma_{RG}$ as the restriction of the syndrome $\sigma$ to the restricted lattice $\mathcal{L}_{RG}$.
We observe that the total number of $R$ and $G$ excitations has to be even, i.e., $|\sigma_{RG}| \equiv 0 \mod 2$, since the excitations in the 2D color code are either created in pairs of the same color, or in triples of three different colors.
Thus, we can always find a subset of edges $\rho_{RG}\subseteq \face 1 {\mathcal{L}_{RG}}$ of $\mathcal{L}_{RG}$, whose $0$-boundary matches the $R$ and $G$ excitations, i.e., $\partial_1 \rho_{RG} = \sigma_{RG}$; see Fig.~\ref{fig_colorcode_2D}(c).
Note that a similar discussion follows for the set of all the $R$ and $B$ excitations $\sigma_{RB}\subseteq\face 0 {\mathcal{L}_{RB}}$.

The first step of the Restriction Decoder is to use the restricted lattices $\mathcal{L}_{RG}$ and $\mathcal{L}_{RB}$ and the restricted syndromes $\sigma_{RG}$ and $\sigma_{RB}$ in order to find $\rho_{RG}\subseteq\face 0 {\mathcal{L}_{RG}}$ and $\rho_{RB}\subseteq\face 0 {\mathcal{L}_{RB}}$ satisfying $\partial_1 \rho_{RG} = \sigma_{RG}$ and $\partial_1 \rho_{RB} = \sigma_{RB}$, respectively.
This step can be viewed as an instance of the MWPM problem, and thus can be efficiently solved by e.g. the blossom algorithm~\cite{Edmonds1965}.
We remark that the task of finding $\rho_{RG}$ and $\rho_{RB}$ is analogous to the problem of decoding the 2D toric code defined on the restricted lattices $\mathcal{L}_{RG}$ and $\mathcal{L}_{RB}$ \cite{Bombin2012a,Delfosse2014}.
As we explain later in Sec.~\ref{sec_decoding}, this observation is a key to generalizing the Restriction Decoder to $d\geq 2$ dimensions.

The second step of the Restriction Decoder uses a local lifting procedure, which allows us to find a correction $\phi$ from $\rho_{RG}\cup\rho_{RB}$.
Let $\beta\subseteq\face 1 {\mathcal{L}}$ be a subset of edges of the lattice $\mathcal{L}$.
We denote by $\facex R 0 \beta$ the set of all the $R$ vertices belonging to the edges of $\beta$ and define $\beta\rest v$ to be the set of all the edges of $\beta$ incident to a vertex $v\in\face 0 {\mathcal{L}}$.
Also, we define $\star 2 v$ to be the set of all the faces of $\mathcal{L}$ containing $v$.
Then, for every vertex $v\in\facex R 0 {\rho_{RG}\cup\rho_{RB}}$, the local lifting procedure finds a subset of faces $\tau_v\subseteq \star 2 v$, such that in the neighborhood of $v$ the $1$-boundary of $\tau_v$ matches $\rho_{RG}\cup\rho_{RB}$, i.e., $(\partial_2 \tau_v)\rest v = (\rho_{RG}\cup\rho_{RB})\rest v$; see Fig.~\ref{fig_colorcode_2D}(d).
As we show in the (Lift) Lemma~\ref{lemma_lift}, there always exists $\tau_v$ satisfying the aforementioned condition.
Moreover, one can find $\tau_v$ in time $\mathcal O(|\star 2 v|)$~\cite{Kubicathesis}.
As claimed, this step of the Restriction Decoder uses only geometrically local information, i.e., $\star 2 v$ and $(\rho_{RG}\cup\rho_{RB})\rest v$.

Finally, the Restriction Decoder returns $\phi = \bigcup_{v\in\facex R 0 {\rho_{RG}\cup\rho_{RB}}} \tau_v$ as the output.
We remark that it is a priori unclear that the output $\phi$ is a valid color code correction, i.e., all the excitations in $\sigma$ are removed by $\phi$.
We address this issue in the (Successful Decoding) Theorem~\ref{thm_success}.
Moreover, the performance of the Restriction Decoder depends on the chosen pair of restricted lattices.

We summarize the Restriction Decoder for the 2D color code on the lattice $\mathcal{L}$.
\begin{itemize}
\item Construct restricted lattices $\mathcal{L}_{RG}$ and $\mathcal{L}_{RB}$.
\item Find restricted syndromes $\sigma_{RG}\subseteq\face 0 {\mathcal{L}_{RG}}$ and $\sigma_{RB}\subseteq\face 0 {\mathcal{L}_{RB}}$.
\item Find $\rho_{RG}\subseteq\face 1 {\mathcal{L}_{RG}}$ and $\rho_{RB}\subseteq\face 1 {\mathcal{L}_{RB}}$ satisfying $\partial_1 \rho_{RG} = \sigma_{RG}$ and $\partial_1 \rho_{RB} = \sigma_{RB}$.
\item For every vertex $v\in\facex R 0 {\rho_{RG}\cup \rho_{RB}}$ find $\tau_v\subseteq\star 2 v$ satisfying $(\partial_2 \tau_v )\rest v = (\rho_{RG}\cup \rho_{RB})\rest v$.
\item Return the correction $\phi = \bigcup_{v\in\facex R 0 {\rho_{RG}\cup\rho_{RB}}} \tau_v$.
\end{itemize}

We emphasize that the Restriction Decoder differs from the decoding algorithm in Ref.~\cite{Delfosse2014} in a couple of important ways; see \cite{Kubicathesis} for a detailed comparison.
First, the Restriction Decoder uses only two restricted lattices $\mathcal{L}_{RG}$ and $\mathcal{L}_{RB}$.
This in turn leads to a substantial improvement of the threshold for the 2D color code on the square-octagon lattice over the thresholds of all the previous efficient color code decoders, as we demonstrate in the next subsection.
Second, the Restriction Decoder uses a local lifting procedure, compared with a global lifting procedure for the decoding algorithm in Ref.~\cite{Delfosse2014}.
This crucial simplification allows us to construct for the first time fully-local decoders for the color code in $d\geq 3$ dimensions.
Third, the Restriction Decoder never aborts and always returns a valid color code correction $\bigcup_{v\in\facex R 0 {\rho_{RG}\cup\rho_{RB}}}\tau_v$.
Fourth, a generalization of the decoding algorithm in Ref.~\cite{Delfosse2014} to $d\geq 2$ dimensions results in a decoder whose complexity would rapidly increase with $d$, as could be seen in three dimensions~\cite{Aloshious2016}.
On the other hand, the Restriction Decoder can be straightforwardly adapted to the $d$-dimensional color code, leading to a simple decoder with quite a satisfactory performance and the time complexity proportional to the time complexity of the toric code decoder that we choose to use.

\subsection{Numerical estimates of the color code threshold}
\label{sec_sqoct}

\begin{figure}[ht!]
\centering
(a)\includegraphics[height = .22\textheight]{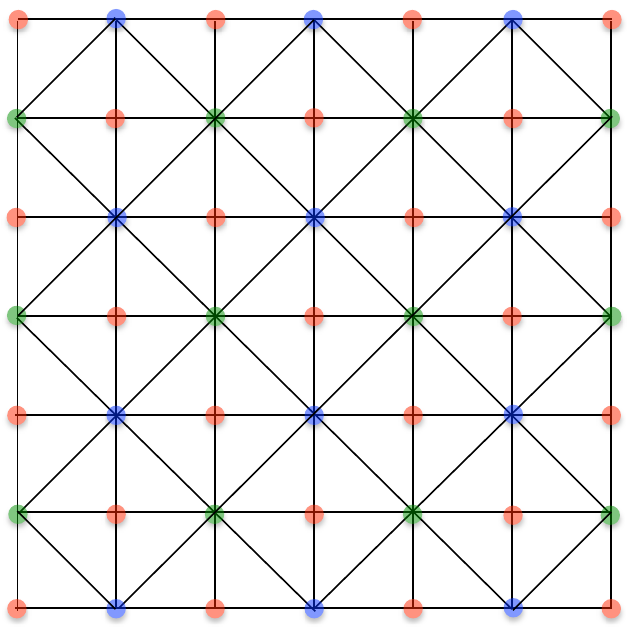}
\hspace*{20mm}
(b)\includegraphics[width=0.4\textwidth,height = .22\textheight]{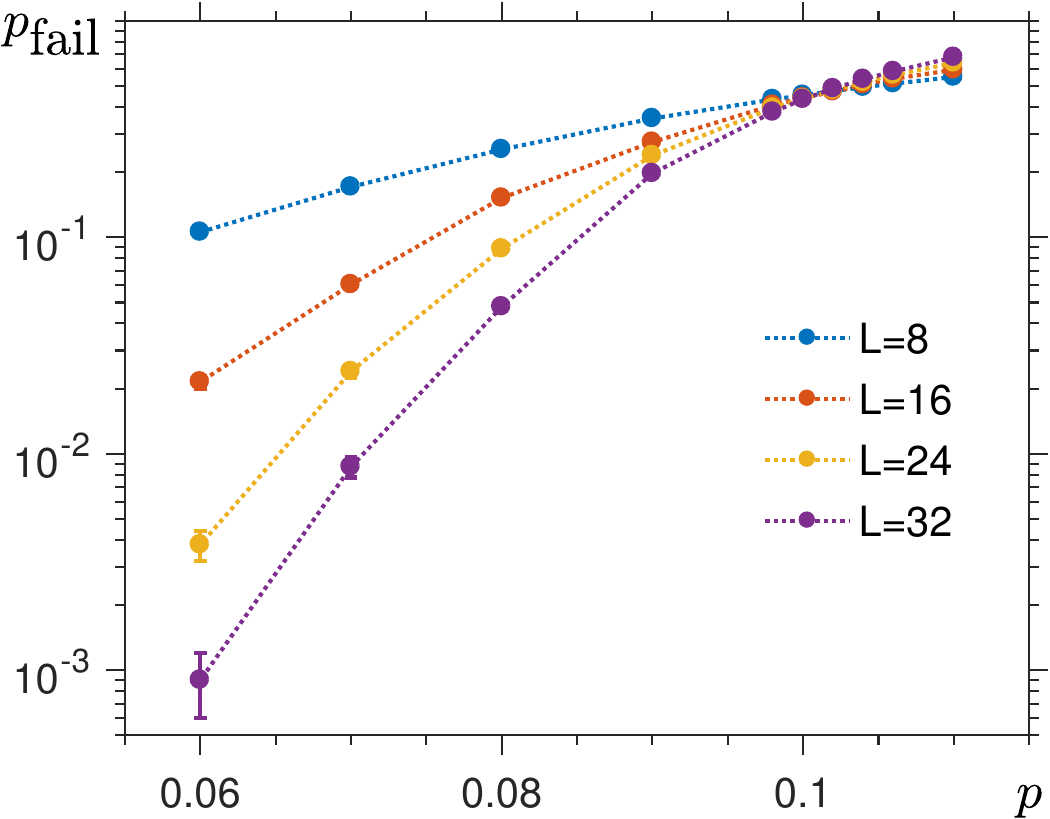}\qquad
\caption{
(a) A two-dimensional color code lattice $\mathcal{L}$ of the linear size $L=6$ with periodic boundary conditions.
We refer to $\mathcal{L}$ as (a lattice dual to) the square-octagon lattice.
(b) The decoding failure probability $p_{\textrm{fail}}(p,L)$ as a function of the phase-flip error rate $p$ and the linear size $L$ of the lattice $\mathcal{L}$.
We estimate the Restriction Decoder threshold $p_{\textrm{2D}} \approx \psqoct$ for the 2D color code on $\mathcal{L}$ by plotting $p_{\textrm{fail}}(p,L)$ for different $L$ and finding their crossing point.
}
\label{fig_sqoct}
\end{figure}

We numerically estimate the Restriction Decoder threshold for the 2D color code and the phase-flip noise, assuming perfect syndrome measurements.
We consider the 2D color code on (a lattice dual to) the square-octagon lattice $\mathcal{L}$ with periodic boundary conditions, as illustrated in Fig.~\ref{fig_sqoct}(a).
We choose the restricted lattices $\mathcal{L}_{RG}$ and $\mathcal{L}_{RB}$ to contain degree-four vertices.
In the first step of the Restriction Decoder we use the Blossom V algorithm provided by Kolmogorov~\cite{Kolmogorov2009} to find $\rho_{RG}$ and $\rho_{RB}$ within the restricted lattices $\mathcal{L}_{RG}$ and $\mathcal{L}_{RB}$.
We estimate the Restriction Decoder threshold $p_{\textrm{2D}} \approx \psqoct$ by plotting the decoding failure probability $p_{\textrm{fail}}(p,L)$ as a function of the phase-flip error rate $p$ for lattices of different linear size $L$ and finding their crossing point; see Fig.~\ref{fig_sqoct}(b).
We report that the 2D color code threshold matches the MWPM algorithm threshold for the 2D toric code on the square lattice.
Lastly, we remark that the choice of restricted lattices determines the Restriction Decoder threshold, as we explain in Sec.~\ref{sec_threshold_estimate}.

\section{Going beyond two dimensions}
\label{sec_basics}

In this overview section, we describe decoding of CSS stabilizer codes.
We start by constructing a CSS chain complex for any CSS code~\cite{Delfosse2014,Bravyi2013b}.
The formalism of chain complexes is useful not only for decoding, but also exemplifies the intimate connection between the CSS codes and the systematic procedure of gauging and ungauging stabilizer symmetries \cite{Kubica2018}.
Then, we discuss geometric lattices which are needed to introduce the toric and color codes in $d\geq 2$ dimensions.
We also briefly explain the problem of toric code decoding.

\subsection{CSS stabilizer codes and chain complexes} 

Any CSS code, whose stabilizer group, by definition, is generated by $X$- and $Z$-type stabilizer generators, can be described by a CSS chain complex
\begin{equation} 
\begin{array}{ccccc}
C_Z & \xrightarrow{\partial_Z} & C_Q & \xrightarrow{\partial_X} & C_X\\
Z\textrm{-stabilizers} & & \mathrm{qubits} & & X\textrm{-stabilizers}
\end{array}
\label{eq_chain}
\end{equation}
where $C_Z$, $C_Q$ and $C_X$ are $\mathbb{F}_2$-linear vector spaces with bases $\mathcal{B}_Z$ = $Z$-type stabilizer generators, $\mathcal{B}_Q$ = physical qubits and $\mathcal{B}_X$ = $X$-type stabilizer generators, respectively.
Note that the vectors in $C_Q$ are in a one-to-one correspondence with the subsets of qubits; similarly, vectors in $C_Z$ and $C_X$ correspond to $Z$- and $X$-type stabilizers.
We choose linear maps $\partial_Z: C_Z \rightarrow C_Q$ and $\partial_X : C_Q \rightarrow C_X$, called the boundary operators, in such a way that
\begin{itemize}
\item the support of any $Z$-type stabilizer $\omega\in C_Z$ is given by $\partial_Z \omega$,
\item the $X$-type syndrome, which is the set of violated $X$-type stabilizer generators for any given $Z$-type error on the subset of qubits $\epsilon\in C_Q$, can be found as $\partial_X \epsilon$.
\end{itemize}
The boundary operators $\partial_Z$ and $\partial_X$ can be identified with the parity-check matrices $H^T_Z$ and $H_X$ of the CSS code.
Note that the composition of two boundary operators is the zero operator, i.e., $\partial_X\circ\partial_Z = 0$, since any $Z$-type stabilizer has a trivial $X$-type syndrome.
We illustrate the CSS chain complex construction with the examples of the toric and color codes in Sec.~\ref{sec_toric_code}~and~Sec.~\ref{sec_color_code}.

Let $\epsilon\in C_Q$ be the subset of physical qubits affected by $Z$ errors.
To decode $Z$ errors we need to guess from the observed $X$-type syndrome $\partial_X\epsilon$ which subset of physical qubits $\varphi\in C_Q$ might have been affected.
Decoding succeeds iff the error $\epsilon$ and the guess $\varphi$ differ by some stabilizer, namely there exists $\omega\in C_Z$ such that $\epsilon + \varphi = \partial_Z \omega$.
Note that the $X$-syndrome of any valid $Z$-correction has to match the $X$-syndrome of the actual $Z$-error, i.e., $\partial_X \varphi = \partial_X \epsilon$.
However, it may happen that the error combined with the correction implements a non-trivial logical $Z$ operator, i.e., $\epsilon+\varphi \not\in\im \partial_Z$, resulting in the decoding failure.
Finally, we remark that $X$ errors can be decoded in a similar way, however one needs to consider a chain complex dual to the one in Eq.~(\ref{eq_chain}).

\subsection{Lattices}

In order to discuss topological stabilizer codes, we need to introduce some notions to succinctly describe geometric lattices on which those codes are defined.
A $d$-dimensional lattice $\mathcal{L}$ can be constructed by attaching $d$-dimensional cells to one another along their $(d-1)$-dimensional faces.
We denote by $\face k {\mathcal{L}}$ the set of all $k$-cells of the lattice $\mathcal{L}$, where $0\leq k\leq d$, and assume that $\face k {\mathcal{L}}$ is finite.
In particular, the sets $\face 0 {\mathcal{L}}$, $\face 1 {\mathcal{L}}$, $\face 2 {\mathcal{L}}$ and $\face 3 {\mathcal{L}}$ correspond to vertices, edges, two-dimensional faces and three-dimensional volumes of $\mathcal{L}$.
We say that the lattice $\mathcal{L}$ is a homogeneous simplicial $d$-complex~\cite{Hatcher2002} if: (i) every $k$-cell of $\mathcal{L}$ forms a $k$-simplex, which together with all its faces belongs to some $d$-cell of $\mathcal{L}$, and (ii) the intersection of any two cells of $\mathcal{L}$ is a face of both of them, where $0\leq k \leq d$.
In what follows, we restrict out attention to lattices without boundary.

\begin{figure}[ht!]
\centering
\includegraphics[width = .9\textwidth]{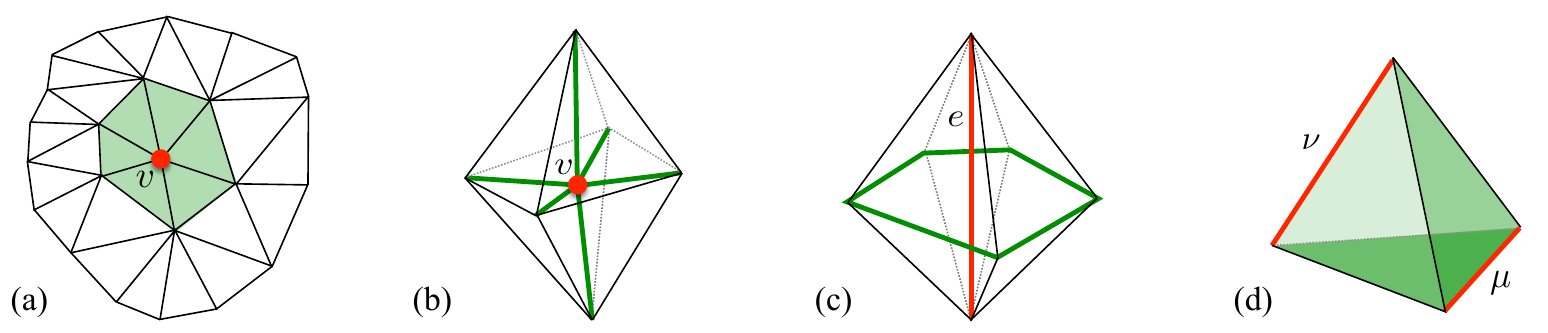}
\caption{
(a) The $2$-star $\star 2 v$ and (b) the $1$-star $\star 1 v$ of the vertex $v$ (red), which respectively correspond to six triangular faces and six edges (shaded in green) containing $v$.
(c) The $1$-link $\link 1 e$ of the edge $e$ (red) contains five edges (green), each of which belongs to the same tetrahedron as $e$ but does not intersect with $e$.
(d) The join $\nu * \mu$ is the tetrahedron (shaded in green) spanned of two edges $\nu$ and $\mu$ (red).
}
\label{fig_starandlink} 
\end{figure}

Now let us analyze the local structure of a lattice $\mathcal L$, which is a homogeneous simplicial $d$-complex.
Since we consider the lattice $\mathcal{L}$ without boundary, every $(d-1)$-simplex of $\mathcal{L}$ belongs to exactly two different $d$-simplices.
Let $\kappa\in\face{k}{\mathcal{L}}$ be a $k$-simplex, where $0\leq k\leq d$.
We abuse the notation and denote by $\face l \kappa$ the set of all $l$-simplices contained in $\kappa$, where $l\leq k$.
The set of all $n$-simplices of $\mathcal{L}$ which contain $\kappa$ is called the $n$-star of $\kappa$ and we denote it by
\begin{equation}
\star{n}{\kappa} = \{ \nu\in\face{n}{\mathcal{L}} | \kappa\in\face{k}{\nu} \},
\end{equation}
where $n\geq k$.
The set of all $n$-simplices of $\mathcal{L}$ which do not intersect with $\kappa$ but belong to the same $d$-simplices as $\kappa$ is called the $n$-link of $\kappa$ and we denote it by
\begin{equation}
\link{n}{\kappa} = \{ \nu\in\face{n}{\mathcal{L}} | \nu\cap\kappa = \emptyset \textrm{ and } \exists\delta\in\star{d}{\kappa}: \nu\in\face{n}{\delta} \},
\end{equation}
where $n\leq d - k -1$.
We write $\linkx{\mathcal{L}}n \kappa$ to explicitly state that we consider the $n$-link of $\delta$ within the lattice $\mathcal{L}$. 
We also define a join $\nu * \mu$ to be the simplex spanned by the vertices of two disjoint simplices $\nu$ and $\mu$.
In particular, the join $\nu * \mu$ is the smallest simplex containing both $\nu$ and $\mu$.
We abuse the notation and write $\nu * \alpha = \{ \nu * \kappa | \forall \kappa \in \alpha\}$, where $\alpha\subseteq\face k {\mathcal{L}}$ is some subset of $k$-simplices.
We illustrate the notions of a star, link and join in Fig.~\ref{fig_starandlink}.

We remark that for any $(d-k-1)$-simplex $\delta\in\face {d-k-1}{\mathcal{L}}$ its $k$-link $\link k \delta$ is homeomorphic to a $k$-dimensional sphere~\cite{Glaser1972}, where $0\leq k < d$.
Moreover, one can show that there is a one-to-one correspondence between the elements of the $k$-link $\link k \delta$ and the elements of the $d$-star $\star d \delta$, namely
\begin{equation}
\label{eq_link_star}
\kappa\in\link{k}{\delta} 
\xlongleftrightarrow{\mu = \delta * \kappa}
\mu\in\star{d}{\delta}.
\end{equation}

\subsection{Boundary operators}

Now we introduce $C_k(\mathcal L)$ to be the $\mathbb{F}_2$-linear vector space with the set of $k$-cells $\face{k}{\mathcal{L}}$ of the lattice $\mathcal{L}$ as a basis. 
Note that there is a one-to-one mapping between vectors in $C_k(\mathcal L)$, which are formal sums of $k$-cells, and subsets of $\face{k}{\mathcal{L}}$, and thus we treat them interchangeably.
We define a (standard) boundary map $\partial_k: C_k(\mathcal L) \rightarrow C_{k-1}(\mathcal L)$ as a linear map specified for every basis element $\kappa \in \face k {\mathcal L}$ by
\begin{equation}
\partial_{k} \kappa = \face {k-1} \kappa = \sum_{\nu \in \face{k-1}\kappa} \nu.
\label{eq_bnd_standard}
\end{equation}
For brevity, we refer to vectors in $C_k(\mathcal L)$ and the corresponding subsets of $\face k {\mathcal{L}}$ as $k$-chains for $\mathcal{L}$.
Let us consider a $k$-chain $\alpha$ for $\mathcal{L}$.
If $\partial_k \alpha = 0$, then we call $\alpha$ a $k$-cycle for $\mathcal{L}$ and write $\alpha\in\ker\partial_k$.
We call $\alpha$ a $k$-boundary for $\mathcal{L}$ and write $\alpha\in\im\partial_{k+1}$ if there exists a $(k+1)$-chain $\beta$ for $\mathcal{L}$, such that $\alpha = \partial_{k+1}\beta$.
Note that any $k$-boundary is a $k$-cycle, i.e., $\im\partial_{k+1} \subseteq \ker\partial_k$, since the composition of two consecutive boundary operators is the zero map, i.e., $\partial_{k}\circ \partial_{k+1} = 0$ for $k\geq 1$.

We remark that we need boundary operators to define the toric code in Sec.~\ref{sec_toric_code}.
However, in order to discuss the color code, it is very convenient to generalize a notion of a boundary operator.
Namely, for all $k \neq n$ we introduce $\bnd k n : C_k \rightarrow C_n$ as a linear map defined on every basis element $\kappa\in \face k {\mathcal{L}}$ by
\begin{equation}
\bnd k n \kappa =
\begin{cases}
\face{n}{\kappa} = \sum_{\nu\in\face{n}{\kappa}} \nu \quad\textrm{if}\ \ k>n,\\
\star{n}{\kappa} = \sum_{\nu\in\star{n}{\kappa}} \nu \quad\textrm{if}\ \ k<n.\\
\end{cases}
\end{equation}
In this notation, $\partial_k = \bnd{k}{k-1}$.
Recall that in general the composition of two generalized boundary operators is not the zero operator, i.e., $\bnd n m \circ \bnd k n \neq 0$, unless $k = n+1 = m + 2$ or the lattice $\mathcal{L}$ satisfies some extra combinatorial conditions as in the case of color code lattices; see Sec.~\ref{sec_color_code}.

\subsection{The toric code in $d\geq 2$ dimensions}
\label{sec_toric_code}

One of the most studied examples of topological stabilizer codes is the toric code.
The $d$-dimensional toric code of type $k$ can be defined on a $d$-dimensional lattice $\mathcal{L}$ built of $d$-dimensional cells, where $1\leq k < d$.
We place one qubit at every $k$-cell $\kappa$ in $\mathcal{L}$.
For every $(k-1)$-cell $\mu$ and $(k+1)$-cell $\nu$ we define $X$- and $Z$-stabilizer generators $S_X(\mu)$ and $S_Z(\nu)$ to be the product of either Pauli $X$ or $Z$ operators on qubits in the neighborhood of $\mu$ and $\nu$, namely 
\begin{equation}
S_X(\mu) = \prod_{\kappa\in\star{k}{\mu}} X(\kappa),\quad\quad
S_Z(\nu) = \prod_{\kappa\in\face{k}{\nu}} Z(\kappa).
\end{equation}
Equivalently, the CSS chain complex from Eq.~(\ref{eq_chain}) associated with the toric code is given by
\begin{equation} 
\begin{array}{ccccc}
C_{k+1}(\mathcal{L}) & \xrightarrow{\partial_{k+1}} & C_{k}(\mathcal{L}) & \xrightarrow{\partial_k} & C_{k-1}(\mathcal{L})\\
Z\textrm{-stabilizers} & & \mathrm{qubits} & & X\textrm{-stabilizers}
\end{array}
\label{eq_chain_tc}
\end{equation}
Non-trivial logical $Z$ operators correspond to the $k$-cycles, which are not $k$-boundaries for $\mathcal{L}$, and thus we view them as $k$-dimensional non-contractible objects.
Any $X$-type syndrome corresponds to a $(k-1)$-boundary for $\mathcal{L}$ and we view it as a collection of the $(d-1)$-dimensional excitations.
Lastly, we remark that non-trivial logical $X$ operators and $Z$-type syndromes can be viewed as $(d-k)$- and $(d-k-1)$-dimensional objects in the dual lattice $\mathcal{L}^*$ (which is obtained from $\mathcal{L}$ by replacing every $k$-cell by a $(d-k)$-cell for all $k$).

To illustrate the discussion, we consider the 3D toric code of type $k=1$ on the cubic lattice $\mathcal{L}$, see Fig.~\ref{fig_restricted_lattice}(b).
We place qubits on edges of $\mathcal{L}$, and define $X$- and $Z$-stabilizer generators for every vertex $v$ and face $f$ of $\mathcal{L}$ as the product of Pauli $X$ and $Z$ operators on qubits on edges adjacent to $v$ and $f$, namely
\begin{equation}
S_X(v) = \prod_{e\in\star{1}{v}} X(e),\quad\quad
S_Z(f) = \prod_{e\in\face{1}{f}} Z(e).
\end{equation}
The logical Pauli $X$ and $Z$ operators form 2D sheet-like and 1D string-like objects, whereas $X$- and $Z$-syndromes can be viewed as 0D point-like and 1D loop-like excitations.

\subsection{Toric code decoding}

Now we discuss the problem of decoding $Z$ errors in the $d$-dimensional toric code of type $k$, where $1\leq k < d$.
Let $\epsilon$ be a $k$-chain for $\mathcal{L}$ corresponding to the subset of qubits affected by $Z$ errors.
As we already mentioned, the information available to the decoder is the $X$-type syndrome $\sigma = \partial_k \epsilon$, which is a $(k-1)$-boundary for $\mathcal{L}$.
We want to find a $k$-chain $\varphi$ for $\mathcal{L}$, which is consistent with the observed syndrome, i.e., $\partial_k \varphi = \sigma$.
Recall that decoding succeeds if the initial error $\epsilon$ and the correction $\phi$ implement a trivial logical operator, i.e., $\epsilon + \phi \in \im\partial_{k+1}$.

One decoding strategy is to choose $\varphi$ as the most likely error for the given syndrome.
Assuming that the phase-flip noise affects every qubit with the same probability, this strategy is equivalent to finding $\varphi$ of the minimal size.
In other words, toric code decoding can be reduced to the following problem, which we call the $k$-Minimum-Weight Filling (MWF) problem:
{\emph{for any $(k-1)$-boundary $\sigma\in\im \partial_{k}$ for the lattice $\mathcal{L}$ find the smallest $k$-chain $\phi$ for $\mathcal{L}$ whose boundary $\partial_k\phi$ matches $\sigma$, i.e.,}
\begin{equation}
k\mathrm{-MWF}(\sigma) = \argmin_{\partial_{k} \varphi  = \sigma} |\varphi|.
\end{equation}
Despite the fact that solving the $k$-MWF problem does not constitute the optimal decoding strategy for the toric code (as it returns the most likely error instead of the most likely equivalence class of errors), it usually yields quite satisfactory error-correction thresholds.
We remark that in the special case of $k=1$ the MWF problem corresponds to the MWPM problem, which can be efficiently solved by e.g. the blossom algorithm~\cite{Edmonds1965}.
However, in a general case for $k\not\in\{1,d-1,d\}$ it is not known how to efficiently find a solution to the MWF problem\footnote{
Note that the case for $k=d$ is equivalent to finding ``inside'' and ``outside'' of some region enclosed by a $(d-1)$-boundary for the $d$-dimensional lattice $\mathcal{L}$, which is easy to solve.
In the case of $k=d-1$ Sullivan \cite{Sullivan1990} proposed an efficient algorithm based on the max-flow min-cut theorem.}.

There is an alternative decoding strategy for the $d$-dimensional toric code of type $k\neq 1$, which does not require solving the MWF problem. 
This efficient strategy is based on a recently proposed cellular automaton, the Sweep Rule, which generalizes celebrated Toom's rule~\cite{Toom1980,Bennett1985} to any locally Euclidean lattice; see the accompanying paper~\cite{Kubica2018toom}.
The resulting toric code decoder, the Sweep Decoder, is fully-local, naturally incorporates parallelization and generically exhibits high error-correction threshold, even in the presence of measurement errors.
As we will see in Sec.~\ref{sec_CA_decoder}, the Sweep Decoder allow us to develop for the very first time cellular-automaton decoders for the color code in $d\geq 3$ dimensions.

\subsection{The color code in $d\geq 2$ dimensions}
\label{sec_color_code}

To define the $d$-dimensional color code, we need a $d$-dimensional lattice $\mathcal{L}$, which is a homogeneous simplicial $d$-complex without boundary.
In addition, we require that the vertices of $\mathcal{L}$ are $(d+1)$-colorable, i.e., one can introduce a function
\begin{equation}
\textrm{color}:\face{0}{\mathcal{L}} \rightarrow \mathbb{Z}_{d+1} = \{ 0, 1, \ldots, d \},
\end{equation}
where $\mathbb{Z}_{d+1}$ is a set of $d+1$ colors and any two vertices connected by an edge have different color.
We remark that one can equivalently define a $d$-dimensional color code lattice to be a $d$-colex.
We postpone the detailed discussion of $d$-colexes to Sec.~\ref{sec_colexes}.

The $d$-dimensional color code of type $k$ is constructed by placing one qubit at every $d$-simplex $\delta\in\face d {\mathcal{L}}$, where $1\leq k<d$.
For every $(k-1)$-simplex $\mu$ and $(d-k-1)$-simplex $\nu$ we define $X$- and $Z$-stabilizer generators $S_X(\mu)$ and $S_Z(\nu)$ to be the product of Pauli $X$ and $Z$ operators on qubits adjacent to $\mu$ and $\nu$, namely
\begin{equation}
S_X(\mu) = \prod_{\delta\in\star{d}{\mu}} X(\delta),\quad\quad
S_Z(\nu) = \prod_{\delta\in\star{d}{\nu}} Z(\delta).
\end{equation}
Equivalently, the CSS chain complex from Eq.~(\ref{eq_chain}) associated with the color code is given by
\begin{equation} 
\begin{array}{ccccc}
C_{d-k-1}(\mathcal{L}) & \xrightarrow{\bnd{d-k-1}{d}} & C_d(\mathcal{L}) & \xrightarrow{\bnd{d}{k-1}} & C_{k-1}(\mathcal{L})\\
Z\textrm{-stabilizers} & & \mathrm{qubits} & & X\textrm{-stabilizers}
\end{array}
\label{eq_chain_cc}
\end{equation}
Note that in Eq.~(\ref{eq_chain_cc}), unlike in Eq.~(\ref{eq_chain_tc}), we do need generalized boundary operators.
Non-trivial logical $Z$ operators correspond to the elements of ${\ker\bnd d {k-1} \setminus \im \bnd {d-k-1} d}$ and can be viewed as $k$-dimensional objects, similarly as for the toric code.
Any $X$-type syndrome is identified with some element of $\im\bnd{d}{k-1}$ and thus can be viewed as a collection of $(k-1)$-dimensional excitations.
We defer the detailed description of logical operators of the $d$-dimensional color code to~Sec.~\ref{sec_logical}.

To illustrate the discussion, we consider the 3D color code of type $k=1$ on the bcc lattice $\mathcal{L}$; see Fig.~\ref{fig_restricted_lattice}(a).
Note that the bcc lattice $\mathcal{L}$ is built of tetrahedral volumes and the vertices of $\mathcal{L}$ are $4$-colorable.
We place qubits on tetrahedra of $\mathcal{L}$, and define $X$- and $Z$-stabilizer generators for every vertex $v$ and edge $e$ of $\mathcal{L}$ as the product of Pauli $X$ and $Z$ operators on qubits on tetrahedra adjacent to $v$ and $e$, namely
\begin{equation}
S_X(v) = \prod_{t\in\star{3}{v}} X(t),\quad\quad
S_Z(e) = \prod_{t\in\star{3}{e}} Z(t).
\end{equation}
The logical Pauli $X$- and $Z$-operators form 2D sheet-like  and 1D string-like objects, whereas $X$- and $Z$-syndromes can be viewed as 0D point-like and 1D loop-like excitations.

\section{Colexes and restricted lattices}  
\label{sec_lattices}

In this section we introduce the notions of colorable $d$-balls and $d$-colexes in order to systematically construct any $d$-dimensional color code lattice $\mathcal{L}$, where $d\geq 2$.
Our definition of $d$-colexes can be viewed as an alternative to the one provided in Ref.~\cite{Bombin2007}.
Then, we prove four technical lemmas regarding color code lattices, which are later needed for color code decoding in Sec.~\ref{sec_decoding}.
Lastly, we discuss the construction of restricted lattices and prove the (Isomorphic Homology Groups) Theorem~\ref{thm_homology} relating homology groups of restricted lattices and the underlying color code lattice $\mathcal{L}$.

\subsection{A local structure of colexes}
\label{sec_colexes}

Now we provide two recursive definitions of a colorable $d$-ball and a $d$-colex.

\begin{definition}[Colorable Ball]
A colorable $d$-ball
\begin{equation}
\mathcal{B}_v = \{ v*\delta | \forall \delta\in {\mathcal{L}} \},
\end{equation}
for $d\geq 1$ is defined as the set of $d$-simplices spanned by a vertex $v$ and all the $(d-1)$-simplices of some $(d-1)$-colex $\mathcal{L}$ homeomorphic to a $(d-1)$-sphere.
The colorable $d$-ball $\mathcal{B}_v$ is homeomorphic to a $d$-disk and its boundary $\partial\mathcal{B}_v$ corresponds to $\mathcal{L}$.
A colorable $0$-ball is defined to be a point.
\end{definition}

\begin{definition}[Colex]
A $d$-colex (without boundary)
\begin{equation}
\mathcal{L} = \bigsqcup_{v\in V} \mathcal{B}_v
\end{equation}
for $d\geq 1$ is a finite collection of disjoint colorable $d$-balls $\mathcal{B}_v$ attached together along their $(d-1)$-dimensional boundaries $\partial \mathcal{B}_v$ in such a way that any $(d-1)$-simplex in $\face{d-1}{\mathcal{L}}$ belongs to the boundary of exactly two different colorable $d$-balls and the vertices of $\mathcal{L}$ are $(d+1)$-colorable.
A $0$-colex is defined to be a collection of colorable $0$-balls.
\end{definition}

\begin{figure}[ht!]
\centering
\includegraphics[width = .9\textwidth]{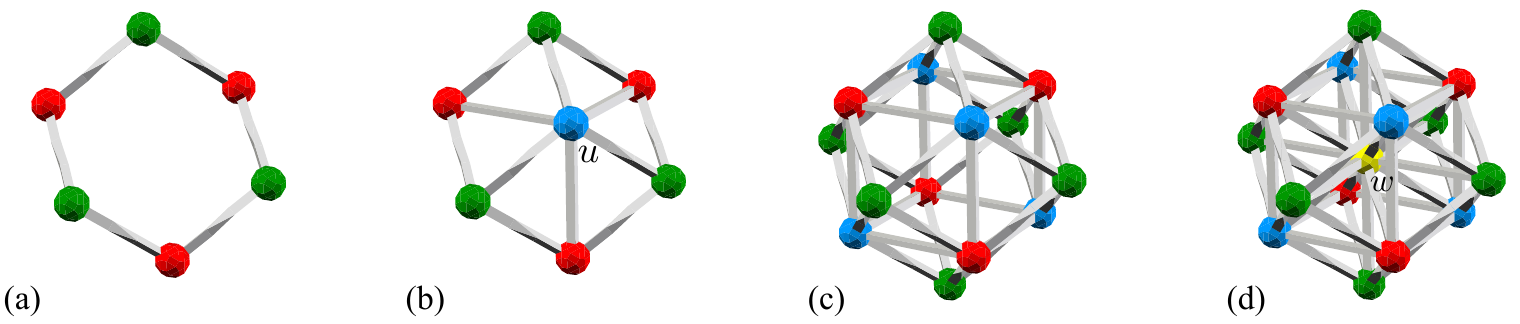}
\caption{
(a) A $1$-colex $\mathcal{L}$ consists of an even number of edges glued together to form a circle.
(b) A colorable $2$-ball $\mathcal{B}_u$ is constructed from the $1$-colex $\mathcal{L}$ by adding triangles spanned by a vertex $u$ and the edges of $\mathcal{L}$.
By definition, $\partial\mathcal{B}_u = \mathcal{L}$.
(c) A $2$-colex ${\mathcal{L}'}$ is a triangulation of a closed two-dimensional manifold, such as a sphere, whose vertices are $3$-colorable.
(d) A colorable $3$-ball $\mathcal{B}_w$ is a collection of tetrahedra spanned by a vertex $w$ and triangular faces of the $2$-colex  ${\mathcal{L}'}$.
By definition, $\partial\mathcal{B}_w = {\mathcal{L}'}$.
The figures were created using vZome available at {\tt{http://vzome.com}}.
}
\label{fig_colexes} 
\end{figure}

Note that a $d$-colex $\mathcal L$ is formally defined as a finite collection of disjoint colorable $d$-balls $\mathcal B_v$, i.e., $\mathcal L = \bigsqcup_{v\in V} \mathcal B_v$, where the vertices in $V$ have the same color $c^*\in\mathbb Z_{d+1}$.
One can show that the colex $\mathcal L$ gives rise to a discretization of some $d$-dimensional closed manifold and, subsequently, for any vertex $u\in\face 0 {\mathcal L}\setminus V$ the set $\star{d}{u}$ satisfies the definition of a colorable $d$-ball.
We thus write $\mathcal B_{u} = \star{d}{u}$.
We leave to the reader to show that for any color $c\in\mathbb Z_{d+1}\setminus \{c^*\}$ the colex $\mathcal L$ can be decomposed as a finite collection of disjoint colorable $d$-balls centered at vertices of $\mathcal L$ of color $c$.

We remark that the $(d+1)$-colorability of the vertices of a $d$-colex $\mathcal{L}$ is sufficient to guarantee that different color code stabilizers commute.
However, this is not a necessary condition to define the color code on $\mathcal{L}$, as just the local colorability within the neighborhood of every vertex of $\mathcal{L}$ is enough to ensure the commutation relations~\cite{Delfosse2013}.
Moreover, the $(d+1)$-colorability of the vertices implies that for any $n<d$ and $d$-simplex $\delta\in\mathcal{L}$, all of the $n$-simplices contained in $\delta$ are of different colors, i.e., if $\mu,\nu\in\face n \delta$ are different, then $\col \nu \neq \col \mu$.
Here, we abuse the notation and write
\begin{equation}
\col \alpha = \bigcup_{v\in\face 0 \alpha} \col v
\end{equation}
to denote the set of colors of all the vertices $v\in\face 0 \alpha$ of an $n$-chain $\alpha$ for $\mathcal{L}$ and refer to it as the color of $\alpha$. 
Lastly, for notational convenience we denote by
\begin{equation}
\facex {c^*} 0 \alpha = \{ v \in\face 0 \alpha | \col v = c^* \}
\end{equation}
the set of all the vertices of $\alpha$ of some chosen color $c^*\in\mathbb{Z}_{d+1}$.

\subsection{Four technical lemmas}

We now present four technical lemmas.
To make them easier to digest, we illustrate each of them with a figure in two dimensions.
Our first lemma is illustrated in Fig.~\ref{fig_technical_lemma1}.

\begin{figure}[ht!]
\centering
\includegraphics[width = .30\textwidth]{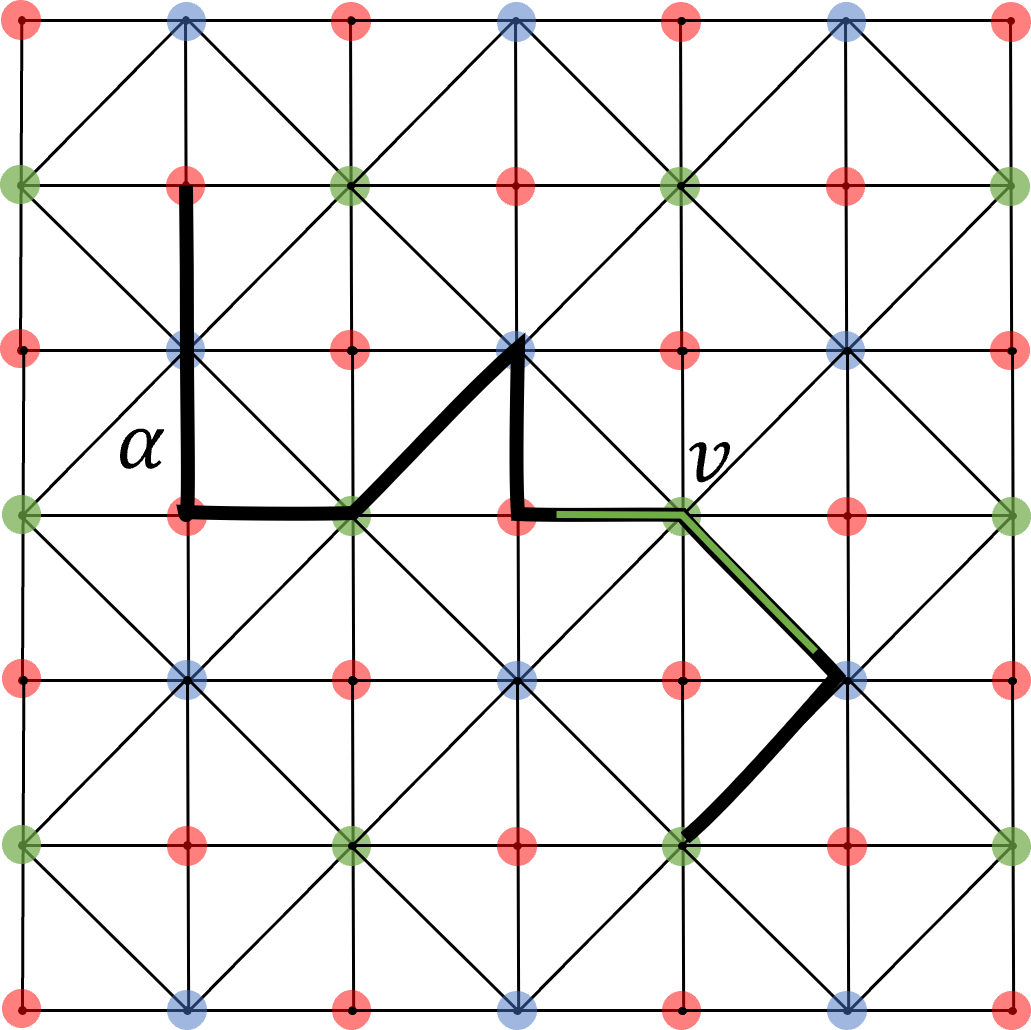}
\hspace{2cm}
\includegraphics[width = .30\textwidth]{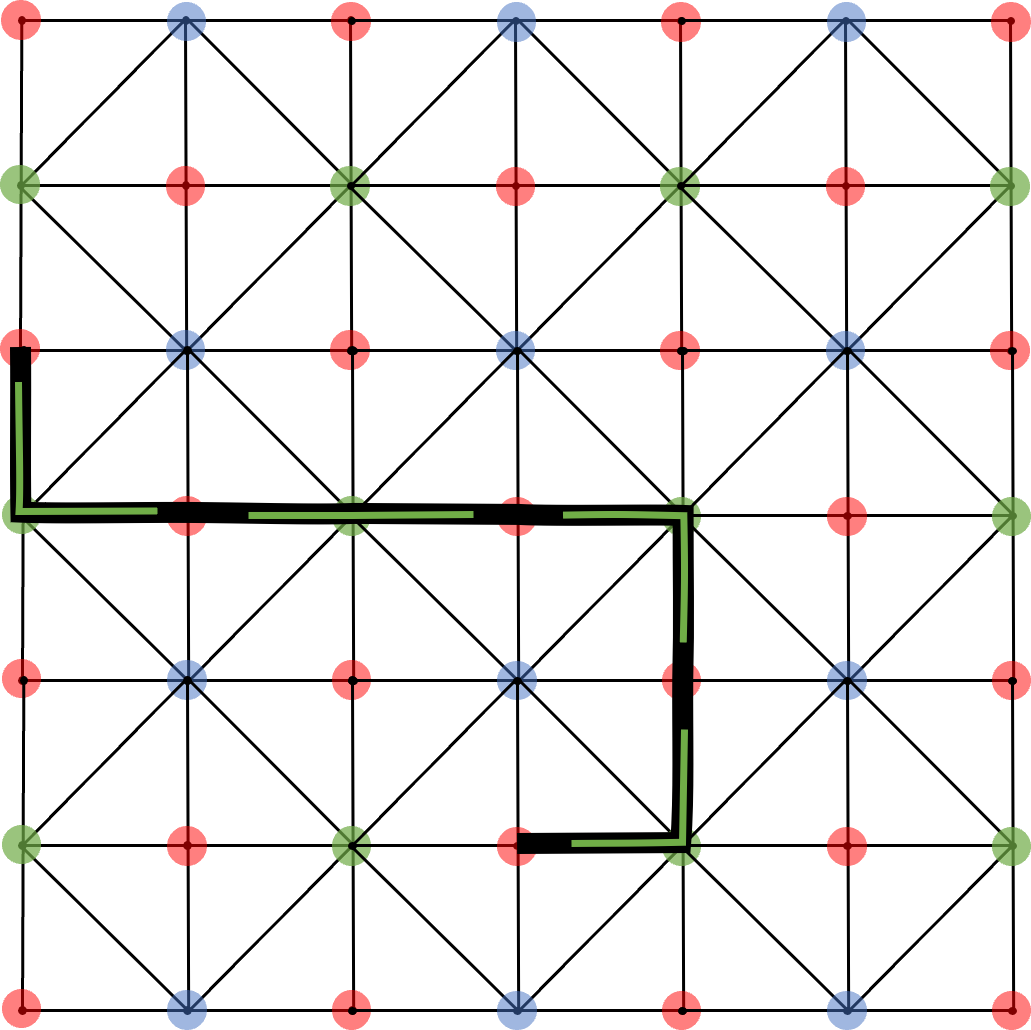}\\
(a) \hspace{7cm} (b)
\caption{
(a) The restriction $\alpha\rest v$ (green) of a 1-chain $\alpha$ (black) in 2D, as defined in Eq.~\eqref{eq_restriction}.
(b) The (Local Restriction) Lemma~\ref{lemma_rest} allows us to partition any 1-chain whose boundary contains a single color into components $\alpha\rest v$ (green) as follows $\alpha = \sum_{v\in \facex {c^*} 0 \alpha} \alpha\rest v$.
}
\label{fig_technical_lemma1} 
\end{figure}

\begin{lemma}[Local Restriction]
\label{lemma_rest}
Let the local restriction of an $n$-chain $\alpha$ for a $d$-colex $\mathcal{L}$ be defined as the set of all the $n$-simplices of $\alpha$ incident to a vertex $v\in\face 0 {\mathcal{L}}$, namely
\begin{equation}
\label{eq_restriction}
\alpha\rest v = \alpha \cap \star n v.
\end{equation}
Let $c^*\in\mathbb{Z}_{d+1}$ be any color.
Then,
\begin{itemize}
\item[(i)] $\alpha\rest u \cap \alpha\rest v = \emptyset$ for any two different $u,v\in \facex {c^*} 0 \alpha$,
\item[(ii)] $\alpha = \alpha' + \sum_{v\in \facex {c^*} 0 \alpha} \alpha\rest v$, where $\col{\alpha'}\subseteq \col\alpha \setminus \{ c^*\}$,
\item[(iii)] $(\partial_n\alpha)\rest v = (\partial_n \alpha\rest v)\rest v$ for any $v\in\face 0 {\mathcal{L}}$,
\item[(iv)] if $v \in\face 0 {\mathcal{L}}\setminus\face 0 {\partial_n\alpha}$, then $\alpha\rest v = v* \partial_n \alpha\rest v$ and $\partial_n \alpha\rest v$ is an $(n-1)$-boundary for a $(d-1)$-colex $\partial\mathcal{B}_v$.
\end{itemize}
\end{lemma}

We would like to clarify that in the (Local Restriction) Lemma~\ref{lemma_rest} and in the rest of the article the operation of taking the local restriction has the highest order of precedence.
For instance, $\partial_n \alpha\rest v$ is unambiguous and stands for $\partial_n (\alpha\rest v)$.

\begin{proof}
Note that (i) and (ii) are an immediate consequence of the $(d+1)$-colorability of the $d$-colex $\mathcal{L}$ and the definition of the local restriction $\alpha\rest v$ in Eq.~(\ref{eq_restriction}).
Moreover, $\alpha + \alpha\rest v$ does not contain any $n$-simplices incident to the vertex $v$, which in turn implies that no $(n-1)$-simplex incident to $v$ belongs to ${\partial_n(\alpha + \alpha\rest v)}$.
By linearity of the boundary map, we thus conclude that an $(n-1)$-simplex $\mu\in\star {n-1} v$ belongs to $\partial_n \alpha$ if and only if it belongs to $\alpha\rest v$, which is the statement of (iii).

To show (iv), observe that $v\not\in\face 0 {\partial_n\alpha\rest v}$ implies $\partial_n \alpha\rest v \subseteq \face {n-1}{\partial\mathcal{B}_v} = \link {n-1} v$.
One can thus show that $\partial_n \alpha\rest v$ is an $(n-1)$-cycle for $\partial\mathcal{B}_v$.
By definition, $\partial\mathcal{B}_v$ is a $(d-1)$-colex homeomorphic to an $(n-1)$-sphere, and hence $\partial_n \alpha\rest v$ is also an $(n-1)$-boundary for $\partial\mathcal{B}_v$.
Lastly, note that for any $n$-simplex $\nu\in\star n v$ there exists a unique $(n-1)$-simplex $\mu\in\link {n-1} v$, such that $\nu = v * \mu$.
Thus, an $(n-1)$-simplex $\mu$ belongs to $\partial_n \alpha\rest v$ if and only if an $n$-simplex $v * \mu$ belongs to $\alpha\rest v$, leading to $\alpha\rest v = v * \partial_n \alpha\rest v$.
This concludes the proof.
\end{proof}

We remark that the following is an immediate corollary of the (Local Restriction) Lemma~\ref{lemma_rest}: if $\alpha$ is an $n$-chain satisfying $|\col\alpha| = n+1$ and $|\col{\partial_n\alpha}| = n$, then $\alpha = \sum_{v\in \facex {c^*} 0 \alpha} \alpha\rest v$ for $\{ c^* \}= \col \alpha \setminus \col{\partial_n\alpha}$.

Fig.~\ref{fig_technical_lemma2} illustrates our second lemma.
\begin{figure}[ht!]
\centering
\includegraphics[width = .75\textwidth]{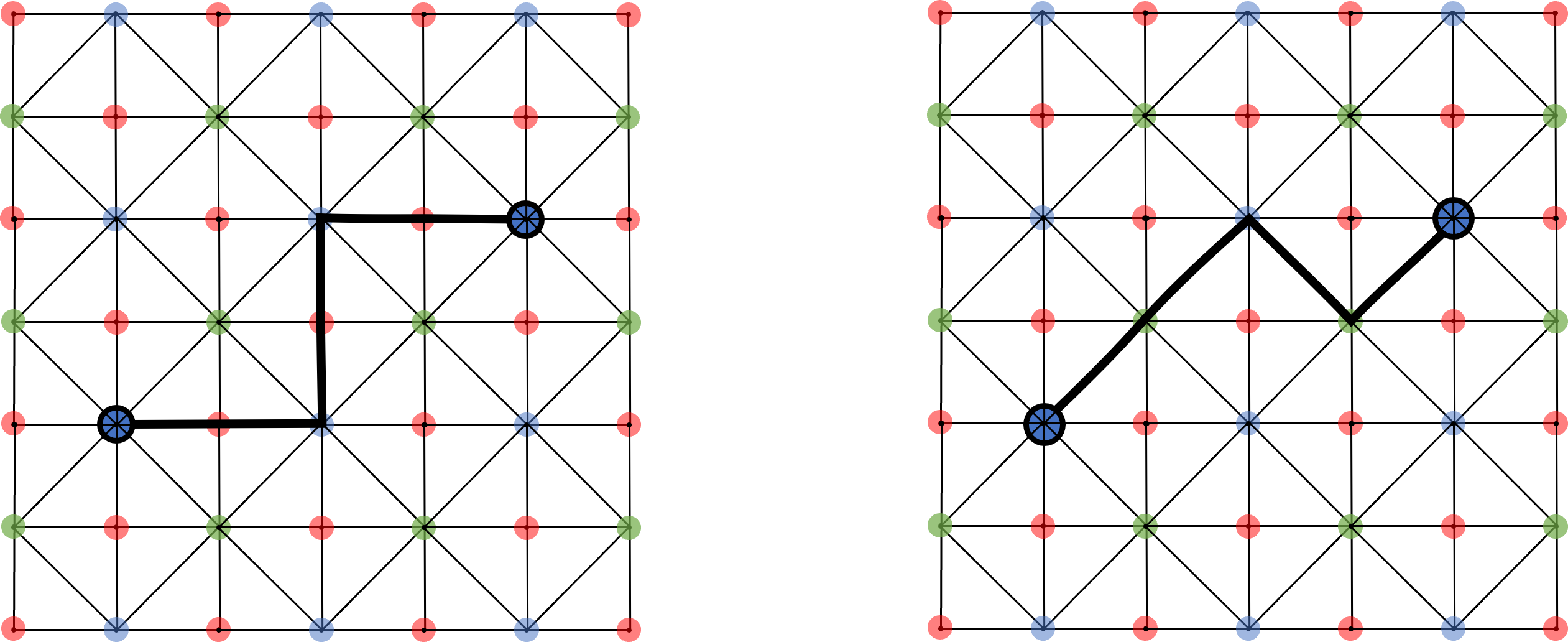}\\
(a) \hspace{7cm} (b)
\caption{
The two $B$ vertices form a $0$-boundary $\alpha$. We can build a 1-chain $\tilde \beta$ of color (a) $RB$ or (b) $GB$ whose boundary is $\alpha$.
}
\label{fig_technical_lemma2} 
\end{figure}

\begin{lemma}[Colorable Filling]
\label{lemma_filling}
Let $\alpha$ be an $n$-boundary for a $d$-colex $\mathcal{L}$, where $n<d$, and $C\subseteq \mathbb{Z}_{d+1}$ be some subset of colors, such that $C\supseteq \col \alpha$ and $|C| > n+1$.
Then, there exists an $(n+1)$-chain $\widetilde\beta$, such that $\partial_{n+1}\widetilde\beta = \alpha$ and $\col {\widetilde\beta} \subseteq C$.
\end{lemma}

\begin{proof}
Since $\alpha$ is an $n$-boundary for $\mathcal{L}$, there exists an $(n+1)$-chain $\beta$ satisfying $\partial_{n+1}\beta = \alpha$.
However, we have no guarantee about the color of $\beta$.
Note that if $n=d-1$, then $|C| > n+1$ implies $C = \mathbb{Z}_{d+1}$, and trivially $\col \beta \subseteq C$.
Thus, it suffices to consider the case of $n<d-1$.
Also, without loss of generality (possibly by reordering the colors) we assume that $C = \mathbb{Z}_{|C|}$.

We proceed with a proof by induction over the colex dimension $d$.
The base case for $d=1$ is immediate, since we then have $n=d-1$.
Now assume the induction hypothesis for any colex of dimension smaller than $d$.
Let $\col {\beta} \subseteq \mathbb{Z}_{i+1}$, where $i\in\col\beta$.
If $i +1 \leq |C|$, then $\widetilde\beta = \beta$ satisfies the assumptions of the lemma.
Otherwise, $i \geq |C|$ and the set of vertices of $\beta$ of color $i$ is non-empty, i.e., $\facex {i} 0 \beta \neq \emptyset$.
Consider an $(n+1)$-chain $\beta\rest v = \beta \cap \star n v$ for the colorable $d$-ball $\mathcal{B}_v$, where $v\in \facex {i} 0 \beta$.
Note that  $v \not\in\face 0 \beta \setminus \face 0 {\partial_{n+1}\beta}$, since $\col v = i$.
Using the (Local Restriction) Lemma~\ref{lemma_rest} we conclude that $\partial_{n+1} \beta\rest v$ is an $n$-boundary for $\partial\mathcal{B}_v$.
By invoking the induction hypothesis for the $(d-1)$-colex $\partial\mathcal B_v$, the $n$-boundary $\partial_{n+1} \beta\rest v$ for $\partial\mathcal{B}_v$ and the subset of colors $\mathbb{Z}_i$, we can find an $(n+1)$-chain $\widetilde\beta(v)$ for $\partial\mathcal{B}_v$, such that 
$\partial_{n+1}\widetilde\beta(v) = \partial_{n+1}(\beta\rest v)$ and $\col{\widetilde\beta(v)} \subseteq \mathbb{Z}_i$.
Therefore, we consider an $(n+1)$-chain $\beta'$ for $\mathcal{L}$ defined as follows
\begin{equation}
\label{eq_betanew}
\beta' = \beta + \sum_{v\in \facex i 0 {\beta}} ( \beta\rest v + \widetilde\beta(v)),
\end{equation}
for which $\partial_{n+1}\beta' = \partial_{n+1}\beta = \alpha$.
Moreover, using the (Local Restriction) Lemma~\ref{lemma_rest} we obtain
\begin{equation}
\col{\beta'} \subseteq \col{\beta + \sum_{v\in \facex i 0 {\beta}} \beta\rest v} \cup \col{ \sum_{v\in \facex i 0 {\beta}} \widetilde\beta(v)}
\subseteq \mathbb{Z}_i.
\end{equation}
Thus, there exists $j<i$, such that $\col{\beta'} \subseteq \mathbb{Z}_{j+1}$ and $j\in\col{\beta'}$.
We repeat the reasoning by redefining $i = j$ and $\beta = \beta'$ until we find $\widetilde\beta$ satisfying the assumptions of the lemma, which concludes the proof.
\end{proof}

An example of our third lemma is shown in Fig.~\ref{fig_technical_lemma3}.
\begin{figure}[ht!]
\centering
\includegraphics[width = .75\textwidth]{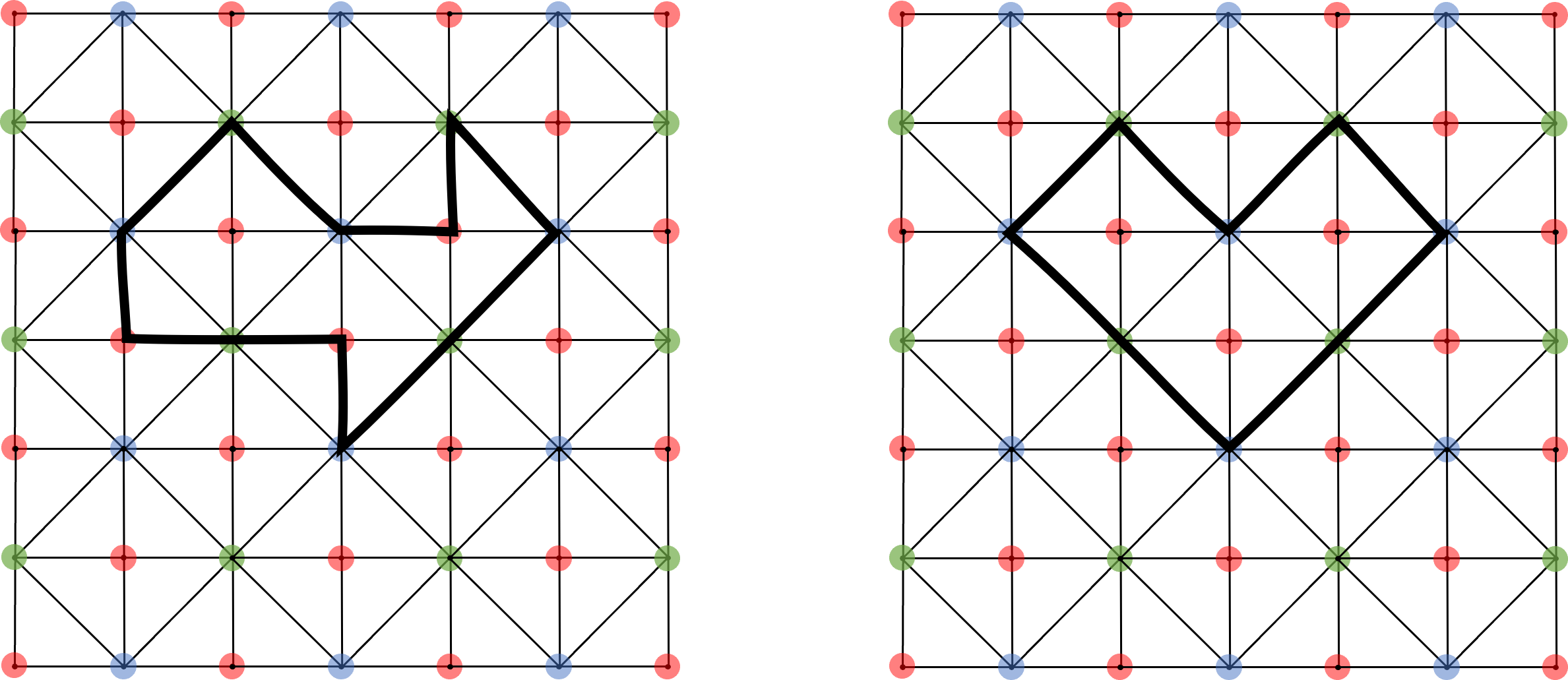}\\
(a) \hspace{7cm} (b)
\caption{
The 1-cycle in (a) which has color $RGB$ is homologous to the 1-cycle in (b) that has color $GB$.
}
\label{fig_technical_lemma3} 
\end{figure}

\begin{lemma}[Colorable Cycles]
\label{lemma_cycles}
Let $\alpha$ be an $n$-cycle for a $d$-colex $\mathcal{L}$, where $1\leq n<d$, and $C\subset \mathbb{Z}_{d+1}$ be any subset of $n+1$ colors.
Then, there exists an $n$-cycle $\widetilde\alpha$ homologous to $\alpha$, i.e., $\widetilde\alpha + \alpha\in\im \partial_{n+1}$, such that $\col {\widetilde\alpha} = C$.
\end{lemma}

\begin{proof}
Without loss of generality, assume that $C = \mathbb{Z}_{n+1}$, $\col\alpha \subseteq \mathbb{Z}_{i+1}$ and $i\in\col\alpha$.
Note that since $\alpha$ is a subset of $n$-simplices of $\mathcal{L}$, thus $|\col\alpha| \geq n+1$.
If $i = n$, then $\widetilde\alpha = \alpha$ satisfies the color requirement.
In the case of $i > n$ the set of vertices of $\alpha$ of color $i$ is non-empty, i.e., $\facex {i} 0 \alpha \neq \emptyset$.
Then, by invoking the (Local Restriction) Lemma~\ref{lemma_rest} we obtain that for every $v \in \facex {i} 0 \alpha$ an $(n-1)$-chain $\partial_n \alpha\rest v$ is an $(n-1)$-boundary for the $(d-1)$-colex $\partial\mathcal{B}_v$ and
it also satisfies $\col{\partial_n \alpha\rest v}\subseteq \mathbb{Z}_i$.
Thus, the (Colorable Filling) Lemma~\ref{lemma_filling} guarantees that we can find an $n$-chain $\widetilde\beta(v)$ for $\partial\mathcal{B}_v$, such that $\partial_n \widetilde\beta(v) = \partial_n \alpha\rest v$ and $\col{\widetilde\beta(v)} \subseteq \mathbb{Z}_i$.
Moreover, $\widetilde\beta(v)$ is homologous to $\alpha\rest v$, since, by definition, $\mathcal{B}_v$ is homeomorphic to a $d$-disk.
Now, we define an $n$-chain $\widetilde\alpha$ for $\mathcal{L}$ as follows
\begin{equation}
\widetilde\alpha = \alpha + \sum_{v\in \facex {i} 0 \alpha} (\alpha\rest v + \widetilde\beta(v)).
\end{equation}
By construction, $\widetilde\alpha$ is homologous to $\alpha$ and from the (Restriction) Lemma~\ref{lemma_rest} we obtain $\col{\widetilde\alpha} \subseteq \mathbb{Z}_i$.
Thus, there exists $j < i$, such that $\col{\widetilde\alpha}\subseteq \mathbb{Z}_{j+1}$ and $j\in \col{\widetilde\alpha}$.
If $j=n$ then $\widetilde\alpha$ satisfies the color requirement, otherwise we repeat the reasoning with $\alpha = \widetilde \alpha$ and $i = j$ until we find $\widetilde\alpha$ of color $C$.
This concludes the proof.
\end{proof}

Fig.~\ref{fig_technical_lemma4} provides an example of our fourth lemma.
\begin{figure}[ht!]
\centering
\includegraphics[width = .34\textwidth]{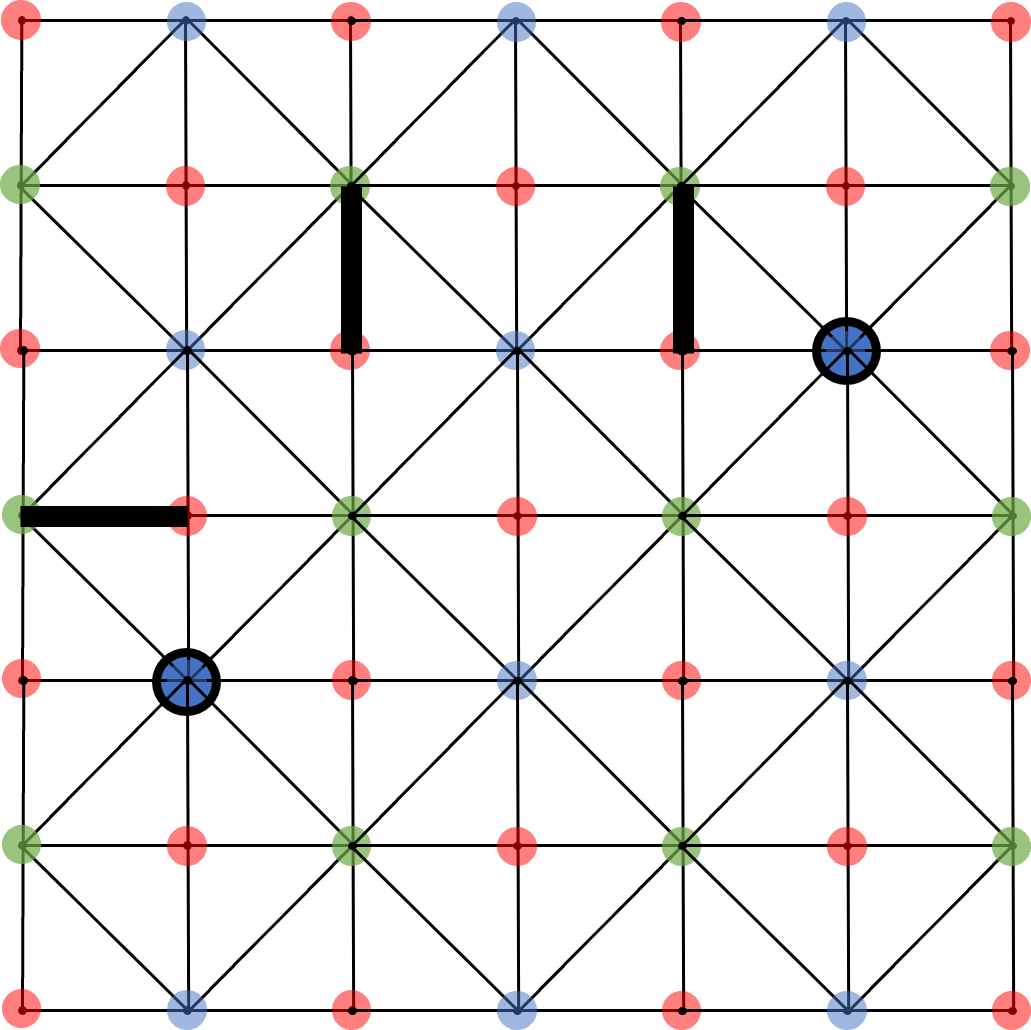}
\caption{
The two highlighted $B$ vertices form a 0-boundary which can be written as the sum of the 0-links of the three $RG$ edges (black).
}
\label{fig_technical_lemma4} 
\end{figure}

\begin{lemma}[Colorable Link]
\label{lemma_link}
Let $\alpha$ be an $n$-boundary for a $d$-colex $\mathcal{L}$, where $n<d$.
Assume that $\alpha$ has $n+1$ colors, i.e., $|\col \alpha| = n+1$.
Then, there exists a $(d-n-1)$-chain $\Omega(\alpha)$ for $\mathcal{L}$, such that $\col{\Omega(\alpha)} = \mathbb{Z}_{d+1}\setminus \col\alpha$ and $\sum_{\mu \in \Omega(\alpha)} \link n \mu = \alpha$.
\end{lemma}

\begin{proof}
Let $c^*\in\mathbb Z_{d+1}\setminus\col\alpha$ and define $C = \col\alpha \sqcup \{ c^* \}$.
The (Colorable Filling) Lemma~\ref{lemma_filling} guarantees the existence of an $(n+1)$-chain $\widetilde\beta$, such that $\partial_{n+1}\widetilde\beta = \alpha$ and $\col{\widetilde\beta} \subseteq C$.
Note that $|\col{\widetilde\beta}| \geq n+2 = |C|$, and thus $\col{\widetilde\beta} = C$.
Thus, the (Local Restriction) Lemma~\ref{lemma_rest} guarantees that $\widetilde\beta$ can be decomposed as
\begin{equation}
\widetilde\beta = \sum_{v \in \facex {c^*} 0 {\widetilde\beta}} \widetilde\beta\rest v,
\end{equation}
and $\partial_{n+1}\widetilde\beta\rest v \subseteq \face n {\partial\mathcal{B}_v}$ for every $v \in \facex {c^*} 0 {\widetilde\beta}$, since $c^* \not\in\col\alpha$.
Moreover, $\partial_{n+1}\widetilde\beta\rest v$ is an $n$-boundary for $\partial\mathcal{B}_v$ and $\partial_{n+1} \widetilde\beta\rest v = \col\alpha$.

If $n = d-1$, then for every $v \in \facex {c^*} 0 {\widetilde\beta}$ the condition $\partial_{n+1}\widetilde\beta\rest v \subseteq \face n {\partial\mathcal{B}_v}$ implies $\widetilde\beta\rest v = \mathcal{B}_v$, and, by definition of a colorable $d$-ball, we get $\partial_{n+1}\widetilde\beta\rest v = \partial \mathcal{B}_v = \link n v$.
Subsequently, by setting $\Omega(\alpha) = \facex {c^*} 0 {\widetilde\beta}$ we find a $(d-n-1)$-chain for $\mathcal{L}$ with the required properties, as $\col{\Omega(\alpha)} = c^*$ and $\alpha = \sum_{v \in \facex {c^*} 0 {\widetilde\beta}} \partial_{n+1} \widetilde\beta\rest v = \sum_{\mu\in\Omega(\alpha)} \link n \mu$.
Thus, it suffices to consider the case of $n<d-1$.

Similarly as in the proof of the (Colorable Filling) Lemma~\ref{lemma_filling}, we proceed with a proof by induction over the colex dimension $d$.
The base case for $d=1$ is immediate, since we then have $n=d-1$.
Now assume the induction hypothesis for any colex of dimension smaller than $d$.
By invoking the induction hypothesis for the $(d-1)$-colex $\partial \mathcal{B}_v$, the $n$-boundary $\partial_{n+1} \widetilde\beta\rest v$ for $\partial \mathcal{B}_v$ and the subset of colors $\partial_{n+1} \widetilde\beta\rest v$, we can find a $(d-n-2)$-chain $\Omega(\partial_{n+1} \widetilde\beta\rest v)$ for $\partial\mathcal{B}_v$, such that 
$\col{\Omega(\partial_{n+1} \widetilde\beta\rest v)} = \col{\partial\mathcal{B}_v} \setminus \col\alpha$ and
\begin{equation}
\partial_{n+1} \widetilde\beta\rest v = \sum_{\mu\in\Omega(\partial_{n+1} \widetilde\beta\rest v)} \linkx {\partial\mathcal{B}_v} n \mu.
\end{equation}
Recall that $\linkx {\partial\mathcal{B}_v} n \mu$ denotes the $n$-link of $\mu\in\face {d-n-2}{\partial\mathcal{B}_v}$ within the lattice $\partial\mathcal{B}_v$ (not within the lattice $\mathcal{L}$).

Now we show that a $(d-n-1)$-chain $\Omega(\alpha)$ for $\mathcal{L}$ defined as follows
\begin{equation}
\Omega(\alpha) = \sum_{v \in \facex {c^*} 0 {\widetilde\beta}} v* \Omega(\partial_{n+1} \widetilde\beta\rest v)
= \sum_{v \in \facex {c^*} 0 {\widetilde\beta}} \sum_{\mu \in \Omega(\partial_{n+1} \widetilde\beta\rest v)} v * \mu
\end{equation}
satisfies the requirements of the lemma.
By construction, we have $\col{\Omega(\alpha)} = \mathbb{Z}_{d+1} \setminus \col\alpha$.
Moreover, one can show that $\linkx {\partial\mathcal{B}_v} n \mu = \link n {v * \mu}$ for all $v\in \facex {c^*} 0 {\widetilde\beta}$ and $\mu\in\Omega(\partial_{n+1}\widetilde\beta\rest v)$. Thus,
\begin{equation}
\sum_{\nu \in \Omega(\alpha)} \link n \nu
= \sum_{v \in \facex {c^*} 0 {\widetilde\beta}} \sum_{\mu \in \Omega(\partial_{n+1} \widetilde\beta\rest v)} \link n {v * \mu}
= \sum_{v \in \facex {c^*} 0 {\widetilde\beta}} \partial_{n+1} \widetilde\beta\rest v = \partial_{n+1} \widetilde\beta = \alpha,
\end{equation}
which concludes the proof.
\end{proof}

\subsection{Restricted lattices}
\label{sec_restricted_lattice}

\begin{definition}[Restricted Lattice]
Let $\mathcal{L}$ be a $d$-colex (without boundary) and $C \subset \mathbb{Z}_{d+1}$ be a subset of $k+1$ colors, where $1\leq k< d$.
The restricted lattice $\mathcal{L}_C$ is a cell $(k+1)$-complex constructed from $\mathcal{L}$ as follows.
\begin{itemize}
\item For $0\leq i \leq k$: $i$-cells in $\mathcal{L}_C$ are the same as $i$-simplices in $\mathcal{L}$ of color included in $C$
\begin{equation}
\face i {\mathcal{L}_C} = \{ \iota\in\face i {\mathcal{L}} | \col\iota \subseteq C\}.
\end{equation}
\item Every $(k+1)$-cell $\Xi(\delta)$ in $\mathcal{L}_C$ is obtained by attaching a $(k+1)$-disk along its boundary to the $k$-link $\link k \delta$ of a $(d-k-1)$-simplex $\delta$ in $\mathcal{L}$ of color $\mathbb{Z}_{d+1}\setminus C$, and thus is uniquely identified with $\delta$
\begin{equation}
\label{eq_new_cells}
\face {k+1}{\mathcal{L}_C} \ni \Xi(\delta) \xlongleftrightarrow{} 
\delta\in\face {d-k-1}{\mathcal{L}},\textrm{ such that } \col{\delta} = {\mathbb{Z}_{d+1}\setminus C}.
\end{equation}
\end{itemize}
\end{definition}

Similarly as for the lattice $\mathcal{L}$, for $i=0,\ldots,k+1$ we introduce $\mathbb{F}_2$-linear vector spaces $C_i(\mathcal{L}_C)$ associated with the $i$-cells of the the restricted lattice $\mathcal{L}_C$ and linear boundary maps
\begin{equation}
\partial_i^C: C_i(\mathcal{L}_C)\rightarrow C_{i-1}(\mathcal{L}_C).
\end{equation}
Note that for $i=1,\ldots, k$ and any $\iota\in\face i {\mathcal{L}_C}$ we have $\partial_i^C \iota = \partial_i \iota$, whereas for any $\Xi(\delta)\in C_{k+1}(\mathcal{L}_C)$, by definition of $(k+1)$-cells of $\mathcal{L}_C$, we have $\partial_{k+1}^C\Xi(\delta)= \link k \delta$.
Moreover, one can show that $\link k \delta$ is a $k$-boundary for $\mathcal{L}$, which in turn leads to $\im\partial_{k+1}^C\subset \im \partial_{k+1}$.

We emphasize that in order to define the toric code of type $k$ on the restricted lattice $\mathcal{L}_C$ we just need to have cells of dimension up to $k+1$ in $\mathcal{L}_C$.
Moreover, the local modifications of the $d$-colex $\mathcal{L}$ we implement to construct $\mathcal{L}_C$ do not change the structure of $k$-cycles and $k$-boundaries for $\mathcal{L}$ (which in turn determine the toric code logical subspace).
We prove this fact in the (Isomorphic Homology Groups) Theorem~\ref{thm_homology}.

\begin{theorem}[Isomorphic Homology Groups]
\label{thm_homology}
The $k^{\textrm{th}}$ homology groups of $\mathcal{L}_C$ and $\mathcal{L}$ are isomorphic, i.e.,
$\ker \partial_{k}^C / \im \partial_{k+1}^C \simeq \ker \partial_{k} / \im \partial_{k+1}$, where $C\subset\mathbb Z_{d+1}$ is a subset of $k+1$ colors and $1\leq k<d$.
\end{theorem}

\begin{proof}
We use the second isomorphism theorem:
\emph{if $U$ and $V$ are submodules of a module $W$, then the quotient modules $(U+V)/V$ and $U/(U\cap V)$ are isomorphic}; see e.g.~\cite{Lang2002}.
Namely, consider the following vector spaces $W = \ker\partial_k$, $U = \ker\partial^C_k$ and $V = \im\partial_{k+1}$.
We have $\ker \partial_k^C \subset \ker\partial_k$ and $\im \partial_{k+1} \subseteq \ker\partial_k$, as any $k$-cycle for $\mathcal{L}_C$ is also a $k$-cycle for $\mathcal{L}$ and any $k$-boundary for $\mathcal{L}$ is a $k$-cycle for $\mathcal{L}$.
In the remaining we prove that: (i) $U + V = \ker\partial_k$ and (ii) $ U \cap V = \im\partial^C_{k+1} $.

To show (i), first note that $U + V \subseteq \ker\partial_k$, since both $U$ and $V$ are subspaces of $W$.
From the (Colorable Cycles) Lemma~\ref{lemma_cycles} we obtain that for any $\alpha\in\ker\partial_k$ we can find $\widetilde\alpha\in\ker\partial^C_{k}$ and $\beta\in\im\partial_{k+1}$ such that $\alpha = \widetilde\alpha + \beta$.
This implies $\ker\partial_k \subseteq U + V$ and we obtain (i).

To show (ii), first note that $\im\partial_{k+1}^C \subseteq U\cap V$, since $\im\partial_{k+1}^C\subset \im\partial_{k+1}$ and $\im\partial_{k+1}^C\subseteq \ker\partial_{k}^C$.
Also, if $\alpha\in\ker\partial^C_k\cap \im\partial_{k+1}$, then $\alpha$ is a $k$-boundary for $\mathcal{L}$ of color $C$.
Thus, using the (Colorable Link) Lemma~\ref{lemma_link} we can find a $(d-k-1)$-chain $\Omega(\alpha)$ for $\mathcal{L}$, such that $\col{\Omega(\alpha)} = \mathbb{Z}_{d+1}\setminus C$ and $\sum_{\mu \in \Omega(\alpha)} \link n \mu = \alpha$.
Note that according to Eq.~(\ref{eq_new_cells}) in the definition of the restricted lattice $\Omega(\alpha)$ corresponds to a $(k+1)$-chain for $\mathcal{L}_C$, namely
$\Xi(\Omega(\alpha)) = \sum_{\mu \in\Omega(\alpha)} \Xi(\mu)$.
Since $\partial_{k+1}^C \Xi(\Omega(\alpha)) = \sum_{\mu \in\Omega(\alpha)} \link k \mu = \alpha$, this implies that $\alpha\in\im\partial_{k+1}^C$ is a $k$-boundary for $\mathcal{L}_C$.
Thus, we conclude that $U\cap V \subseteq \im\partial_{k+1}^C$ and obtain (ii).
We finish the proof by invoking the second isomorphism theorem and concluding that $\ker \partial_{k}^C / \im \partial_{k+1}^C \simeq \ker \partial_{k} / \im \partial_{k+1}$.
\end{proof}

Lastly we remark that we can successively add $i$-cells to the restricted lattice $\mathcal{L}_C$ to construct a $d$-dimensional lattice $\widetilde{\mathcal{L}_C}$, where $k+2\leq i \leq d$.
Namely, every $i$-cell of $\widetilde{\mathcal{L}_C}$ corresponds to some $(d-i)$-simplex $\delta\in\face {d-i}{\mathcal{L}}$ of color in $\mathbb{Z}_{d+1} \setminus C$ and is constructed by first finding all $(d-i+1)$-simplices of color in $\mathbb{Z}_{d+1}\setminus C$ that contain $\delta$, and then attaching an $i$-disk along its boundary to the corresponding $(i-1)$-disks forming an $(i-1)$-dimensional sphere (which have already been attached in the preceding inductive step).
This inductive construction of $\widetilde{\mathcal{L_C}}$ can be viewed as a description of the $i$-skeleton of a cell complex corresponding to $\widetilde{\mathcal{L_C}}$.
We thus can treat $\widetilde{\mathcal{L}_C}$ on the same footing as $\mathcal{L}$, namely as a discretization of some $d$-dimensional manifold.

\section{Decoding the color code in $d\geq 2$ dimensions}  
\label{sec_decoding}

This section is devoted to the problem of decoding the $d$-dimensional color code, where $d\geq 2$.
First, we establish a morphism of chain complexes of the $d$-dimensional color and toric codes, which is a generalization of the result for $d=2$ from Ref.~\cite{Delfosse2014}.
Then, we discuss the structure of logical Pauli operators of the color code.
Finally, we introduce the Restriction Decoder for the $d$-dimensional color code and prove the (Successful Decoding) Theorem~\ref{thm_success} describing the performance of the Restriction Decoder. 
This in turn allows us to estimate the color code threshold from the toric code threshold.

\subsection{Morphism between color and toric code chain complexes}

\begin{definition}[Restriction]
Let $\mathcal{L}$ be a $d$-colex and $C\subset \mathbb{Z}_{d+1}$ be a subset of $k+1$ colors, where $1\leq k < d$.
The restriction $\pi_C$ is a triple of linear operators
$\pi^{(0)}_C : C_{k-1}({\mathcal L}) \rightarrow C_{k-1}({\mathcal L}_C)$,
${\pi^{(1)}_C : C_{d}({\mathcal L})\rightarrow C_{k}({\mathcal L}_C)}$ and
$\pi^{(2)}_C : C_{d-k-1}({\mathcal L}) \rightarrow C_{k+1}({\mathcal L}_C)$
defined as follows
\begin{eqnarray}
\pi^{(0)}_C(\mu) &=& 
\begin{cases}
\mu &\mathrm{\ if \ } \col{\mu} \subset C, \\
0 &\mathrm{\ otherwise},
\end{cases}\\
\pi^{(1)}_C(\delta) &=&  \delta_C,\\
\pi^{(2)}_C(\nu) &=& 
\begin{cases}
\Xi(\nu) &\mathrm{\ if \ } \col{\nu}= \mathbb{Z}_{d+1}\setminus C,\\
0 &\mathrm{\ otherwise},
\end{cases}
\end{eqnarray}
where $\delta_C$ is defined as the $k$-simplex of color $C$ belonging to the $d$-simplex $\delta\in\face{d}{\mathcal{L}}$.
Recall that $\mathcal{L}_C$ is the restricted lattice and, according to Eq.~(\ref{eq_new_cells}), $\Xi(\nu)$ is the $(k+1)$-face of ${\mathcal L}_C$ corresponding to the $(d-k-1)$-simplex $\nu$ removed from $\mathcal{L}$, and thus, by definition, $\partial_{k+1}^C \Xi(\alpha) = \link k \nu$.
\end{definition}

\begin{lemma}[Morphism]
\label{lemma_morphism}
Consider the color code of type $k$ defined on a $d$-colex $\mathcal{L}$, where $1\leq k < d$ and let $C \subset \mathbb{Z}_{d+1}$ be a subset of $k+1$ colors.
Then, the restriction $\pi_C$ is a morphism between chain complexes of the color code of type $k$ on $\mathcal{L}$ and the toric code of type $k$ on the restricted lattice $\mathcal{L}_C$.
In other words, the following diagram is commutative
\begin{equation}
\begin{CD}
C_{d-k-1}({\mathcal L}) @> \bnd{d-k-1}{d} >>
C_{d}({\mathcal L}) @> \bnd{d}{k-1} >>
C_{k-1}({\mathcal L})\\
@VV \pi^{(2)}_C V @VV \pi^{(1)}_C V @VV \pi^{(0)}_C V\\
C_{k+1}({\mathcal L}_C) @> \partial_{k+1}^C >> 
C_{k}({\mathcal L}_C) @> \partial_{k}^C >>
C_{k-1}({\mathcal L}_C)\\
\end{CD}
\label{eq_diagram}
\end{equation}
\end{lemma}

\begin{proof}

Let us pick $\delta\in\mathcal{L}$ and consider the right side of the diagram.
We want to show that $\pi^{(0)}_C \circ \bnd{d}{k-1} (\delta) = \partial_{k}^C  \circ \pi^{(1)}_C (\delta)$.
Note that for any $n<k$ all $n$-simplices of $\delta$ of colors included in $C$ belong to the $k$-simplex $\delta_C = \pi_C^{(1)}(\delta)$ of $\delta$, namely
\begin{equation}
\face{n}{\delta_C} = \{ \nu \in\face{n}{\delta} | \col{\nu} \subset C\}.
\end{equation}
Thus, we obtain
\begin{eqnarray}
\pi^{(0)}_C \circ \bnd{d}{k-1} (\delta) &=&
\pi^{(0)}_C \left(\sum_{\mu\in\face{k-1}{\delta}} \mu \right) = 
\sum_{\substack{\mu\in\face{k-1}{\delta} \\ \col{\mu}\subset C}} \mu = 
\sum_{\mu\in\face{k-1}{\delta_C}} \mu = \partial_{k} (\delta_C)\\
&=& \partial_{k}^C  \circ \pi^{(1)}_C (\delta),
\end{eqnarray}
which shows commutativity of the right side of the diagram in Eq.~(\ref{eq_diagram}).

Now we analyze the left side of the diagram.
Let us pick $\delta\in\face{d-k-1}{\mathcal{L}}$ and consider two cases.
In the first case, when $\col{\delta} = \mathbb{Z}_{d+1}\setminus C$, all the $k$-simplices in the $k$-link of $\delta$ have color $C$.
Recall that Eq.~(\ref{eq_link_star}) establishes a one-to-one correspondence between the elements of the $k$-link $\link k \delta$ and the elements of the $d$-star $\star d \delta$, which allows us to represent any $\mu\in\star d \delta$ as $\mu = \delta *\kappa$ with $\kappa\in\link{k}{\delta}$.
Then, we also have $\pi^{(1)}_C (\delta * \kappa) = \kappa$ and thus
\begin{eqnarray}
\pi^{(1)}_C \circ \bnd{d-k-1}{d} (\delta)
&=&\pi^{(1)}_C \left(\sum_{\mu\in\star{d}{\delta}} \mu \right)
= \pi^{(1)}_C \left(\sum_{\kappa\in\link{k}{\delta}} \delta * \kappa \right)
= \sum_{\kappa\in\link{k}{\delta}} \kappa\\
&=& \partial_{k+1}^C(\Xi(\delta)) = \partial_{k+1}^C \circ \pi^{(2)}_C (\delta).
\end{eqnarray}
In the other case, i.e., $\col{\delta} \neq \mathbb{Z}_{d+1}\setminus C$, we define
$\overline C = C \cup \col \delta$, whose cardinality satisfies the following
\begin{equation}
|\overline C | = |C| + |\col \delta| - |C \cap \col \delta | \leq (k+1) + (d-k) - 1 = d.
\end{equation}
Let $x = |\overline C | - 1$.
Since $\overline C \supseteq \col \delta$, then the $d$-star $\star d \delta$ of $\delta$ can be decomposed as a disjoint union of $d$-stars of $x$-simplices $\xi$ of color $\overline C$, which contain $\delta$; see the (Disjoint Union) Lemma 2 in Ref.~\cite{Kubica2015a}.
In other words
\begin{equation}
\star{d}{\delta} = \sum_{\substack{\xi\in\star{x}{\delta}\\ \col{\xi} = \overline C}} \star{d}{\xi}.
\end{equation}
Moreover, for any $n$-simplex $\nu$ with $n< d$ the cardinality of the $d$-star $\star d \nu$ is even, i.e., $|\star{d}{\nu}| \equiv 0 \mod 2$; see the (Even Support) Lemma 4 in Ref.~\cite{Kubica2015a}.
Since we consider arithmetic in $\mathbb{F}_2$,  for any $x$-simplex $\xi$ we have $\sum_{\mu\in \star d \xi} \xi_C = 0$, where $\xi_C$ denotes the $k$-simplex of color $C$ belonging to $\xi$.
Thus, we obtain
\begin{eqnarray}
\pi^{(1)}_C \circ \bnd{d-k-1}{d} (\delta) &=& \pi^{(1)}_C \left(\sum_{\mu\in\star{d}{\delta}} \mu \right) =
\pi^{(1)}_C \left(\sum_{\substack{\xi\in\star{x}{\delta} \\ \col{\xi} = \overline C} } \sum_{\mu\in\star{d}{\xi}} \mu \right)\\
&=& 
\sum_{\substack{\xi\in\star{x}{\delta} \\ \col{\xi} = \overline C} } \sum_{\mu\in\star{d}{\xi}} \pi^{(1)}_C (\mu) 
= \sum_{\substack{\xi\in\star{x}{\delta} \\ \col{\xi} = \overline C} } \sum_{\mu\in\star{d}{\xi}} \xi_C\\
&=& 0 = \partial_{k+1}^C \circ \pi^{(2)}_C (\delta),
\end{eqnarray}
where in the last step we use $\pi^{(2)}_C (\delta) = 0$, since $\col \delta \neq \mathbb{Z}_{d+1} \setminus C$.
This shows commutativity of the left side of the diagram in Eq.~(\ref{eq_diagram}) and thus concludes the proof.
\end{proof}

\subsection{Logical operators of the color code}
\label{sec_logical}

\begin{definition}[Membrane]
Let $\mathcal{L}$ be a $d$-colex and $\gamma\in \ker\partial_k / \im\partial_{k+1}$ be an element of the $k^{\textrm{th}}$ homology group of $\mathcal{L}$, where $1\leq k<d$.
Let $C\subset\mathbb{Z}_{d+1}$ be a subset of $k$ colors and $c^*\in\mathbb{Z}_{d+1}\setminus C$.
We define a $(\gamma,C)$-membrane to be a $d$-chain $\lambda_{\gamma,C}$ for $\mathcal{L}$ specified as follows
\begin{equation}
\label{eq_membrane}
\lambda_{\gamma,C} = \sum_{v\in \facex {c^*} 0 {\widetilde\gamma}} \sum_{\mu\in\widetilde\Omega(v)} \star d \mu,
\end{equation}
where $\widetilde\gamma$ is a representative of $\gamma$, such that $\col{\widetilde\gamma} = C \sqcup \{c^*\}$, and
$\widetilde\Omega(v)\subseteq \star {d-k} v$ satisfies $\col{\widetilde\Omega(v)} = \mathbb Z_{d+1}\setminus C$ and
$\sum_{\mu\in\widetilde\Omega(v)} \link {k-1} \mu = \partial_{k}\widetilde\gamma\rest v$.
\end{definition}

We would like to make a couple of remarks on the definition of a $(\gamma,C)$-membrane.
\begin{enumerate}
\item[(i)] The (Colorable Cycles) Lemma~\ref{lemma_cycles} guarantees that we can find an $k$-cycle $\widetilde\gamma$ for $\mathcal{L}$, which is a representative of $\gamma$ and $\col{\widetilde\gamma} = C \sqcup \{c^*\}$.
\item[(ii)] By using the (Local Restriction) Lemma~\ref{lemma_rest} we obtain that $\partial_k\widetilde\gamma\rest v$ is a $(k-1)$-boundary for the $(d-1)$-colex $\partial\mathcal{B}_v$ and $\col{\partial_k\widetilde\gamma\rest v} = C$ for every vertex $v\in\facex{c^*} 0 {\widetilde\gamma}$.
Thus, the (Colorable Link) Lemma~\ref{lemma_link} guarantees the existence of a $(d-k-1)$-chain $\Omega(\partial_k\widetilde\gamma\rest v)$ for $\partial\mathcal{B}_v$, such that
\begin{equation}
\sum_{\mu\in\Omega(\partial_k\widetilde\gamma\rest v)} \linkx {\partial\mathcal{B}_v} {k-1} \mu = \partial_k\widetilde\gamma\rest v.
\end{equation}
Recall that $\linkx {\partial\mathcal{B}_v} {k-1} \mu$ denotes the $(k-1)$-link of $\mu\in\face {d-k-1}{\partial\mathcal{B}_v}$ within the lattice $\partial\mathcal{B}_v$.
{One can show that $\linkx {\partial\mathcal{B}_v} {k-1} \mu = \link {k-1} {v * \mu}$}, and thus a subset $\widetilde\Omega(v) \subseteq\star {d-k} v$ defined as follows
\begin{equation}
\widetilde\Omega(v) = v * \Omega(\partial_k\widetilde\gamma\rest v) = \sum_{\mu\in\Omega(\partial_k\widetilde\gamma\rest v)} v * \mu
\end{equation}
 satisfies $\col{\widetilde\Omega(v)} = \mathbb{Z}_{d+1}\setminus C$ and
$\sum_{\nu\in\widetilde\Omega(v)} \link {k-1} \nu = \partial_{k}\widetilde\gamma\rest v$.
Note that $\Omega(\partial_k\widetilde\gamma\rest v)$ is not guaranteed to be unique, and thus neither is $\widetilde\Omega(v)$.
Since $\partial_{k}\widetilde\gamma\rest v \subseteq \face {k-1}{\partial\mathcal{B}_v}$, by using the (Local Restriction) Lemma~\ref{lemma_rest} we arrive at
\begin{equation}
\label{eq_gammarest}
\widetilde\gamma\rest v =v * \partial_k\widetilde\gamma\rest v
= \sum_{\nu\in\widetilde\Omega(\partial_k\widetilde\gamma\rest v)} v * \link {k-1} \nu.
\end{equation}
\item[(iii)] A $(\gamma,C)$-membrane $\lambda_{\gamma,C}$ is uniquely specified the choice of $\widetilde\gamma$ and $\widetilde\Omega(v)$ for every vertex $v\in\facex {c^*} 0 {\widetilde\gamma}$.
As we will see, different possible $(\gamma,C)$-membranes correspond to different representatives of the same logical operator of the color code.
To provide the reader with some intuition, we remark that green shaded faces in  Fig.~\ref{fig_colorcode_2D}(a) form a $(\gamma,C)$-membrane $\lambda_{\gamma,C}$, where $\gamma$ is associated with a dashed green curve and $C$ denotes color $G$.
We can then set $c^*$ as color $R$ and choose $\widetilde\gamma$ to be the $1$-chain of color $RG$ that borders green shaded faces, i.e., $\widetilde\gamma = \pi^{(1)}_{C^*}(\lambda_{\gamma,C})$, where $C^* = C\sqcup{c^*}$.
Subsequently, $\widetilde\Omega(v)$ includes all the edges of color $RB$ that are incident to $v$ and cross the dashed green curve.
\end{enumerate}

\begin{lemma}[Logical Operators]
Let $\mathcal{L}$ be a $d$-colex for the color code of type $k$ and $C\subset\mathbb{Z}_{d+1}$ be a subset of $k$ colors, where $1\leq k < d$.
Let $\gamma\in\ker\partial_k / \im\partial_{k+1}$ be an element of the $k^{\textrm{th}}$ homology group of $\mathcal{L}$.
Then, a $Z$-type operator supported on a $(\gamma,C)$-membrane is a logical operator for the color code, i.e.,
$\lambda_{\gamma,C}\in\ker\bnd d {k-1}$.
Moreover, if $\gamma \in(\ker\partial_k\setminus\im\partial_{k+1}) / \im\partial_{k+1}$, then the logical operator is non-trivial, i.e.,
$\lambda_{\gamma,C} \not\in\im\bnd {d-k-1} d$.
\end{lemma}

\begin{proof}
We can straightforwardly verify that $\lambda_{\gamma,C}$ specified in Eq.~(\ref{eq_membrane}) is a logical operator for the color code.
Namely,
\begin{eqnarray}
\bnd d {k-1} \lambda_{\gamma,C}
&=& \sum_{\facex {c^*} 0 {\widetilde\gamma}} \sum_{\nu\in\widetilde\Omega(v)} \bnd d {k-1} \star d \nu
= \sum_{\facex {c^*} 0 {\widetilde\gamma}} \sum_{\nu\in\widetilde\Omega(v)} \link {k-1} \nu
= \sum_{\facex {c^*} 0 {\widetilde\gamma}} \partial_k \widetilde\gamma\rest v\\
&=& \partial_k \widetilde\gamma = 0. 
\end{eqnarray}
Now let us assume that $\lambda_{\gamma,C} \in\im\bnd {d-k-1} d$.
Then, from the (Morphism) Lemma~\ref{lemma_morphism} we obtain $\pi^{(1)}_{C^*}(\lambda_{\gamma,C}) \in\im\partial^{C^*}_{k+1}$, where $C^*= C \sqcup \{ c^* \}$.
Note that for any vertex $v\in\facex {c^*} 0 {\widetilde\gamma}$ and $\nu\in\widetilde\Omega(v)$ we have $\col\nu = \mathbb{Z}_{d+1}\setminus C$. 
Since $v* \link {k-1} \nu$ is the set of all the $k$-simplices of $\star d \nu$ with color $C^*$, thus
$\pi^{(1)}_{C^*}(\star d \nu) = v* \link {k-1} \nu$ and
\begin{equation}
\pi^{(1)}_{C^*}(\lambda_{\gamma,C})
= \sum_{\facex {c^*} 0 {\widetilde\gamma}}  \sum_{\nu\in\widetilde\Omega(v)} \pi^{(1)}_{C^*}(\star d \nu)
= \sum_{\facex {c^*} 0 {\widetilde\gamma}} \sum_{\nu\in\widetilde\Omega(v)}  v * \link {k-1} \nu
= \sum_{\facex {c^*} 0 {\widetilde\gamma}} \widetilde\gamma\rest v 
= \widetilde\gamma,
\end{equation}
where we use Eq.~(\ref{eq_gammarest}) and the (Local Restriction) Lemma~\ref{lemma_rest}.
This leads to $\widetilde\gamma\in\im\partial_{k+1}^{C^*} \subset \im\partial_{k+1}$, and by the contrapositive we conclude that if $\gamma\in(\ker\partial_k\setminus\im\partial_{k+1}) / \im\partial_{k+1}$, then $\lambda_{\gamma,C} \not\in\im\bnd {d-k-1} d$, which finishes the proof.
\end{proof}

Now we make a couple of remarks on the logical operators of the color code.
\begin{enumerate}
\item[(i)] Consider a collection of subsets of $k$ colors not containing some chosen color $c^* \in \mathbb{Z}_{d+1}$, i.e.,
\begin{equation}
\label{eq_basis_color}
\mathcal{C} = \{ C\subset\mathbb{Z}_{d+1} | c^* \not\in C \textrm{ and }|C|=k\}
\end{equation}
and a basis of the $k^{\textrm{th}}$ homology group for $\mathcal{L}$, i.e.,
\begin{equation}
\label{eq_basis_cycle}
\Gamma = \{ \gamma_1,\ldots,\gamma_{b_k} \},
\end{equation}
where $b_k = \dim(\ker\partial_k /\im\partial_{k+1})$ is the $k^{\textrm{th}}$ Betti number for $\mathcal{L}$.
Then, one can show that the set of $Z$-type operators supported on $(\gamma,C)$-membranes
$\{ \lambda_{\gamma,C}\} _{\gamma\in \Gamma, C\in\mathcal{C}}$
is a basis of $Z$-type logical operators for the color code on $\mathcal{L}$; see Ref.~\cite{Bombin2007}.
Thus, the number of logical qubits of the color code is equal to $|\Gamma| \cdot |\mathcal{C}| = b_k {d\choose k} $.
\item[(ii)] One can verify that $\pi_{C^*}^{(1)}(\lambda_{\gamma,C}) = \widetilde\gamma$ for all $\gamma\in\Gamma$ and $C\in\mathcal{C}$.
Thus, using the (Morphism) Lemma~\ref{lemma_morphism} and the (Isomorphic Homology Groups) Theorem~\ref{thm_homology} one concludes that $\{ \pi_{C^*}^{(1)}(\lambda_{\gamma,C}) \}_{\gamma\in\Gamma}$ is a basis of $Z$-type logical operators for the toric code on the restricted lattice $\mathcal{L}_{C^*}$, where $C^* = C \sqcup \{ c^* \}$.
This, in turn, leads to the following isomorphism
\begin{equation}
\label{eq_logical_isomorphism}
\ker\bnd d {k-1} / \im \bnd {d-k-1} d \simeq \prod_{C\in\mathcal{C}} \ker\partial_k^{C^*} / \im\partial_{k+1}^{C^*}
\end{equation}
between the logical operators of the color code on $\mathcal{L}$ and the logical operators of $|\mathcal{C}| = {d \choose k}$ different toric codes, each defined on a different restricted lattice $\mathcal{L}_{C^*}$.
\end{enumerate}

\subsection{Restriction Decoder for the color code}
\label{sec_restriction_decoder}

\begin{algorithm}[h]
\caption{Restriction Decoder}
\vspace*{8pt}
\SetKwInOut{Require}{Require}
\Require{$d$-colex $\mathcal{L}$ for the color code of type $k$, where $1\leq k < d$\\
toric code decoder {\tt{TCdecoder}}\\
local lifting procedure {\tt{Lift}}
}
\KwIn{color code syndrome, i.e., a $(k-1)$-chain $\sigma\in\im\bnd{d}{k-1}$ for $\mathcal{L}$}
\vspace*{2pt}
\KwOut{color code correction, i.e., a $d$-chain $\tau$ for $\mathcal{L}$}
\vspace*{8pt}
define $\mathcal{C} = \{ C\subset\mathbb{Z}_{d+1} | c^* \not\in C \textrm{ and }|C|=k\}$ for some chosen color $c^* \in \mathbb{Z}_{d+1}$\\
\vspace*{2pt} 
\For{$C\in\mathcal{C}$}{
\hspace*{5mm} define $C^* = C \sqcup \{ c^* \}$\\
\hspace*{5mm} find the restricted lattice $\mathcal{L}_{C^*}$ and the restricted syndrome $\pi^{(0)}_{C^*}(\sigma)$\\
\hspace*{5mm} $\rho_C = {\texttt{TCdecoder}(\mathcal{L}_{C^*}, \pi^{(0)}_{C^*}(\sigma))}$
\hfill{\emph{// decode $\pi^{(0)}_{C^*}(\sigma)$ for the toric code on $\mathcal{L}_{C^*}$}}
}
define $\rho = \sum_{C\in\mathcal{C}} \rho_C$\\
find $\facex {c^*} 0 \rho = \{ v\in\face 0 \rho | \col v = c^* \}$\\ 
\For{$v\in \facex {c^*} 0 \rho$}{
\hspace*{5mm} find the colorable $d$-ball $\mathcal{B}_v$, and the local restrictions $\rho\rest v$ and $\sigma\rest v$\\
\hspace*{5mm}
$\tau_v = \texttt{Lift}(\mathcal{B}_v,\rho\rest v, \sigma\rest v)$
\hfill{\emph{// locally lift the toric code correction $\rho$}}
}
\vspace*{-2pt}
\KwRet{$\sum_{v\in \facex {c^*} 0 \rho} \tau_v$}
\vspace*{8pt}
\end{algorithm}

We would like to make a couple of remarks on the Restriction Decoder.
\begin{enumerate}

\item[(i)] ${\texttt{TCdecoder}(\mathcal{L}_{C^*}, \pi^{(0)}_{C^*}(\sigma))}$ is a decoder of the $d$-dimensional toric code of type $k$ defined on the lattice $\mathcal{L}_{C^*}$.
The decoder takes as the input the lattice $\mathcal{L}_{C^*}$ and the syndrome $\pi^{(0)}_{C^*}(\sigma) \in\im\partial_k^{C^*}$, and returns some toric code correction, i.e., a $k$-chain $\rho_C$ for $\mathcal{L}_{C^*}$ satisfying $\partial_k^{C^*} \rho_C = \pi^{(0)}_{C^*}(\sigma)$.
We emphasize that one can use as $\texttt{TCdecoder}$ any toric code decoder.

\item[(ii)] $\texttt{Lift}(\mathcal{B}_v,\rho\rest v, \sigma\rest v)$ uses only geometrically local information about the color code lattice $\mathcal{L}$, the combined toric code correction $\rho$ and the color code syndrome $\sigma$.
Namely, as the input $\texttt{Lift}$ takes the colorable $d$-ball $\mathcal{B}_v$, and the local restrictions $\rho\rest v$ and $\sigma\rest v$.
By definition, we require that as the output $\texttt{Lift}$ returns a $d$-chain $\tau_v \subseteq \mathcal{B}_v$ satisfying
\begin{equation}
\label{eq_cond_lift}
\forall C\in\mathcal{C}: \pi^{(1)}_{C^*} (\tau_v) = \rho_C\rest v.
\end{equation}
A priori, it is not clear that one can always find $\tau_v$ satisfying the condition in Eq.~(\ref{eq_cond_lift}) or that different possible choices of $\tau_v$ do not affect whether or not the Restriction Decoder successfully corrects the error $\epsilon$.
We address those issues in the (Lift) Lemma~\ref{lemma_lift} and the (Successful Decoding) Theorem~\ref{thm_success}.

\item[(iii)] The local lifting procedure $\texttt{Lift}$ can be used to construct $(\gamma,C)$-membranes $\lambda_{\gamma,C}$, and thus to find representatives of logical operators for the color code.
Namely,
\begin{equation}
\lambda_{\gamma,C} = \sum_{v\in\facex{c^*} 0 {\widetilde\gamma}} \mathtt{Lift}(\mathcal{B}_v,\widetilde\gamma\rest v, 0),
\end{equation}
where $\widetilde\gamma$ is a representative of $\gamma$ satisfying $\col{\widetilde\gamma} = C \sqcup \{ c^* \}$.

\item[(iv)] The complexity of the Restriction Decoder is determined by the complexity of the toric code decoder $\texttt{TCdecoder}$.
Moreover, as long as $| \star d v |$ is upper-bounded by some constant for all $v\in\face 0 {\mathcal{L}}$, the local lifting procedure $\texttt{Lift}$ can be implemented in constant time (albeit in time exponential in $| \star d v |$).
Namely, one can naively find a $d$-chain $\tau_v$ by considering all possible subsets of $\star d v$ and finding one which satisfies Eq.~(\ref{eq_cond_lift}).
For more efficient ways of implementing $\texttt{Lift}$, see Ref.~\cite{Kubicathesis}.

\item[(v)] Lastly, we emphasize that the Restriction Decoder is fully local (in the sense of space and time) as long as $\texttt{TCdecoder}$ is fully local.
In particular, one obtains a cellular-automaton decoder for the $d$-dimensional color code of type $k$, where $2\leq k < d$, by using as $\texttt{TCdecoder}$ the Sweep Decoder from Ref.~\cite{Kubica2018toom}.
We discuss that in Sec.~\ref{sec_CA_decoder}.
\end{enumerate}

\begin{lemma}[Lift]
\label{lemma_lift}
For every $v\in \facex {c^*} 0 \rho$ there exists a $d$-chain $\tau_v \subseteq \mathcal{B}_v$ satisfying Eq.~(\ref{eq_cond_lift}).
\end{lemma}

\begin{proof}
First note that since $\sigma\in\im\bnd {d} {k-1}$, thus there has to exist a $d$-chain $\widetilde\tau\subseteq\mathcal{B}_v$, such that
\begin{equation}
\label{eq_sometau}
(\bnd d {k-1} \widetilde\tau)\rest v = \sigma\rest v.
\end{equation}
In particular, if $v\not\in\face 0 \sigma$, then we set $\widetilde\tau = 0$.
We can show that for all $C\in\mathcal{C}$ the following $(k-1)$-chain for $\mathcal{B}_v$
\begin{equation}
\xi(C) = \partial_k^{C^*}\left(\rho_C\rest v + \pi^{(1)}_{C^*}(\widetilde\tau)\right)
\end{equation}
is a $(k-1)$-boundary for the $(d-1)$-colex $\partial\mathcal{B}_v$.
Thus, the (Colorable Link) Lemma~\ref{lemma_link} guarantees that for all $C\in\mathcal{C}$ we can find a $(d-k-1)$-chain $\Omega(\xi(C))$ satisfying $\sum_{\mu\in\Omega(\xi(C))} \linkx {\partial\mathcal{B}_v} {k-1} \mu = \partial_k(\rho_C\rest v)$.
Recall that 
$\linkx {\partial\mathcal{B}_v} {k-1} \mu = \link {k-1} {v * \mu}$.
This in turn implies that $\rho_C\rest v = \sum_{\mu\in\Omega(\xi(C))} v*\link {k-1} {v*\mu}$.
One can verify that by setting
\begin{equation}
\tau_v = \widetilde\tau + \sum_{C\in\mathcal{C}} \sum_{\mu\in\Omega(\xi(C))} \star d {v*\mu}
\end{equation}
one finds a $d$-chain $\tau_v$ satisfying Eq.~(\ref{eq_cond_lift}).

In the rest of the proof we show that for all $C\in\mathcal{C}$ the $(k-1)$-chain $\xi(C)$ is indeed a $(k-1)$-boundary for $\partial\mathcal{B}_v$.
First, Eq.~(\ref{eq_sometau}) implies that $\pi^{(0)}_{C^*}((\bnd d {k-1} \widetilde\tau)\rest v) = \pi^{(0)}_{C^*}(\sigma\rest v)$, as the latter is derived from Eq.~(\ref{eq_sometau}) by restricting attention to the $(k-1)$-chains of color within $C^*$.
Thus, using the (Morphism) Lemma~\ref{lemma_morphism} and the fact that the operators $\pi^{(0)}_{C^*}(\cdot)$ and $\cdot\rest v$ commute, we arrive at
\begin{equation}
\label{eq_xxx1}
\left(\partial_k^{C^*}\circ \pi^{(1)}_{C^*}(\widetilde\tau)\right)\rest v
=  \left(\pi^{(0)}_{C^*}\circ \bnd d {k-1} (\widetilde\tau)\right)\rest v
= \pi^{(0)}_{C^*}(( \bnd d {k-1} \widetilde\tau)\rest v)= \pi^{(0)}_{C^*}(\sigma)\rest v.
\end{equation}
By the construction of $\rho_C$ as a valid toric code correction, we have $\partial_k^{C^*}\rho_C = \pi_{C^*}^{(0)}(\sigma)$.
Using the (Local Restriction) Lemma~\ref{lemma_rest} we obtain
\begin{equation}
\label{eq_xxx2}
\pi^{(0)}_{C^*}(\sigma)\rest v = (\partial_k^{C^*} \rho_C)\rest v = (\partial_k^{C^*} \rho_C\rest v)\rest v.
\end{equation}
Combining Eqs.~(\ref{eq_xxx1})~and~(\ref{eq_xxx2}) leads to
\begin{equation}
\partial_k^{C^*}\left(\rho_C\rest v + \pi^{(1)}_C(\widetilde\tau)\right)\rest v = 0.
\end{equation}
Thus, $v\not\in\face 0 {\xi(C)}$ and the (Restriction) Lemma~\ref{lemma_rest} implies that $\xi(C)$ is a $(k-1)$-boundary for $\partial\mathcal{B}_v$, which concludes the proof.
\end{proof}

\begin{theorem}[Successful Decoding]
\label{thm_success}
Let $\mathcal{L}$ be a $d$-colex for the color code of type $k$, where $1\leq k < d$.
Let $\epsilon\subseteq \mathcal{L}$ be a $Z$-type error and $\sigma = \bnd d {k-1}\epsilon$ be the corresponding $X$-syndrome.
Then, the output $\sum_{v\in \facex {c^*} 0 \rho} \tau_v$ of the Restriction Decoder is a valid color code correction for the error $\epsilon$, i.e.,
 $\epsilon + \sum_{v\in \facex {c^*} 0 \rho} \tau_v \in \ker \bnd d {k-1}$.
Moreover, the Restriction Decoder introduces no logical error iff for all $C\in\mathcal{C}$ the toric code decoder
${\mathtt{TCdecoder}(\mathcal{L}_{C^*}, \pi^{(0)}_{C^*}(\sigma))}$ introduces no logical error, i.e.,
\begin{equation}
\label{eq_equivalence}
\epsilon + \sum_{v\in \facex {c^*} 0 \rho} \tau_v \in \im \bnd {d-k-1} d \iff \forall C\in\mathcal{C}: \pi_{C^*}^{(1)}(\epsilon) + \rho_C \in \im\partial_{k+1}^{C^*}.
\end{equation}
\end{theorem}

\begin{proof}
First note that $\bigcup_{C\in\mathcal{C}} C = \mathbb{Z}_{d+1}$.
Thus, one can show that $\epsilon + \sum_{v\in \facex {c^*} 0 \rho} \tau_v \in \ker \bnd d {k-1}$ holds if and only if $\pi^{(0)}_{C^*}\circ\bnd d {k-1}(\epsilon + \sum_{v\in \facex {c^*} 0 \rho} \tau_v) = 0$ for all $C\in\mathcal{C}$.
We straightforwardly verify the latter.
By definition of $\rho_C$ we have $\partial_k^{C^*} \rho_C = \pi^{(0)}_{C^*}(\sigma)$. 
On the other hand, by using the (Local Restriction) Lemma~\ref{lemma_rest} we find
\begin{equation}
\rho_C = \sum_{v\in\facex {c^*} 0 {\rho_C} } \rho_C\rest v = \sum_{v\in \facex {c^*} 0 \rho} \rho_C\rest v.
\end{equation}
Note that in the above we use two facts: (i) $\facex {c^*} 0 {\rho_C} \subseteq \facex {c^*} 0 \rho$ (since $\rho_C \subseteq \rho$) and (ii) $\rho_C\rest v = 0$ for all $v\in \facex {c^*} 0 \rho \setminus \facex {c^*} 0 {\rho_C}$.
This leads to
\begin{equation}
0 = \pi^{(0)}_{C^*}(\sigma) + \partial_k^{C^*} \rho_C = \pi^{(0)}_{C^*}(\sigma) + \sum_{v\in \facex {c^*} 0 \rho} \partial_k^{C^*} (\rho_C\rest v).
\end{equation}
Therefore, by using the the (Morphism) Lemma~\ref{lemma_morphism} and Eq.~(\ref{eq_cond_lift}) we obtain
\begin{eqnarray}
\pi^{(0)}_{C^*}\circ\bnd d {k-1}\left(\epsilon + \sum_{v\in \facex {c^*} 0 \rho} \tau_v\right) 
&=& \pi^{(0)}_{C^*}(\sigma) + \sum_{v\in \facex {c^*} 0 \rho} \partial_k^{C^*}(\pi^{(1)}_{C^*}(\tau_v))\\
&=& \pi^{(0)}_{C^*}(\sigma) + \sum_{v\in \facex {c^*} 0 \rho} \partial_k^{C^*}(\rho_C\rest v) = 0
\end{eqnarray}
for any $C\in\mathcal{C}$.
Thus, the output of the Restriction Decoder is indeed a valid color code correction.

To show the equivalence in Eq.~(\ref{eq_equivalence}) we observe that for all $C\in\mathcal{C}$
\begin{equation}
\label{eq_proj1}
\pi^{(1)}_{C^*}\left(\epsilon + \sum_{v\in \facex {c^*} 0 \rho} \tau_v\right)
= \pi^{(1)}_{C^*}(\epsilon) + \sum_{v\in \facex {c^*} 0 \rho} \rho_C\rest v = \pi^{(1)}_{C^*}(\epsilon) + \rho_C,
\end{equation}
where we use Eq.~(\ref{eq_cond_lift}).
Since $\epsilon + \sum_{v\in \facex {c^*} 0 \rho} \tau_v\in\ker\bnd d {k-1}$ and $\pi^{(1)}_{C^*}(\epsilon) + \rho_C\in\ker\partial_k^{C^*}$, then by invoking the isomorphism in Eq.~(\ref{eq_logical_isomorphism}) we conclude that the equivalence in Eq.~(\ref{eq_equivalence}) has to hold.
This finishes the proof of the theorem.
\end{proof}

\subsection{Estimating the color code threshold from the toric code threshold}
\label{sec_threshold_estimate}

We can find an estimate of the Restriction Decoder threshold for the $d$-dimensional color code of type $k$ on the lattice $\mathcal{L}$ in terms of the thresholds of the $d$-dimensional toric codes of type $k$ on the restricted lattices $\mathcal{L}_{C^*}$ for $C\in\mathcal{C}$.
Recall that $1\leq k <d$, $\mathcal{C} = \{ C \subset\mathbb{Z}_{d+1} | c^* \not\in C \textrm{ and } |C| = k \}$ and $C^* = C \sqcup \{ c^* \}$ for some chosen color $c^*\in\mathbb{Z}_{d+1}$.
Namely, from the (Successful Decoding) Theorem~\ref{thm_success} we know that the Restriction Decoder introduces no logical error if and only if $\mathtt{TCDecoder}$ for the toric code on $\mathcal{L}_{C^*}$ introduces no logical error for all $C\in\mathcal{C}$.
If the noise affecting the color code on $\mathcal{L}$ is modeled by the bit-flip or phase-flip noise, then the effective noise affecting the toric code on $\mathcal{L}_{C^*}$ is also described by the bit-flip or phase-flip noise.
Importantly, no correlations between errors within the same restricted lattice $\mathcal{L}$ are introduced, however the noise strength changes and may vary for different qubits depending on the local structure of $\mathcal{L}_{C^*}$.

Let the color code on $\mathcal{L}$ be affected by the phase-flip noise of strength $p$ and $f_{C^*}(p)$ capture the effective noise strength affecting the toric code on $\mathcal{L}_{C^*}$.
Note that the function $f_{C^*}(\cdot)$ depends on the local structure of $\mathcal{L}_{C^*}$.
Let $q_{C^*}$ be the $\mathtt{TCDecoder}$ threshold for the toric code on $\mathcal{L}_{C^*}$.
By definition, if $f_{C^*}(p) < q_{C^*}$, then with high probability $\mathtt{TCDecoder}$ introduces no logical error.
Note that the performance of the Restriction Decoder is limited by the worst performance of $\mathtt{TCDecoder}$ for the toric codes on the restricted lattices $\mathcal{L}_{C^*}$ for all $C\in\mathcal{C}$.
Thus, the Restriction Decoder threshold $p_{\mathcal{L}}$ for the color code on $\mathcal{L}$ can be estimated as
\begin{equation}
\label{eq_threshold_estimate}
p_{\mathcal{L}} = \min_{C\in\mathcal C } f^{-1}_{C^*} (q_{C^*}),
\end{equation}
where $f^{-1}_{C^*}(\cdot)$ denotes the inverse function of $f_{C^*}(\cdot)$ (in the neighborhood $(0,\varepsilon)$ where it exists).

\section{More examples of decoding in 2D and 3D}
\label{sec_examples}

In this section we illustrate color code decoding using the Restriction Decoder with a couple of examples in 2D and 3D.
In particular, we discuss some variants of the Restriction Decoder, which: (i) provide an almost-linear time decoding, (ii) do not need restricted lattices, and (iii) can be implemented via cellular automata.
Also, we numerically estimate the Restriction Decoder thresholds for the 3D color code from the 3D toric code thresholds.

\subsection{Almost-linear time decoding of 0D point-like excitations in the color code}
\label{sec_UF}

\begin{figure}[ht!]
\centering
(a)\includegraphics[width=0.4\textwidth,height = .22\textheight]{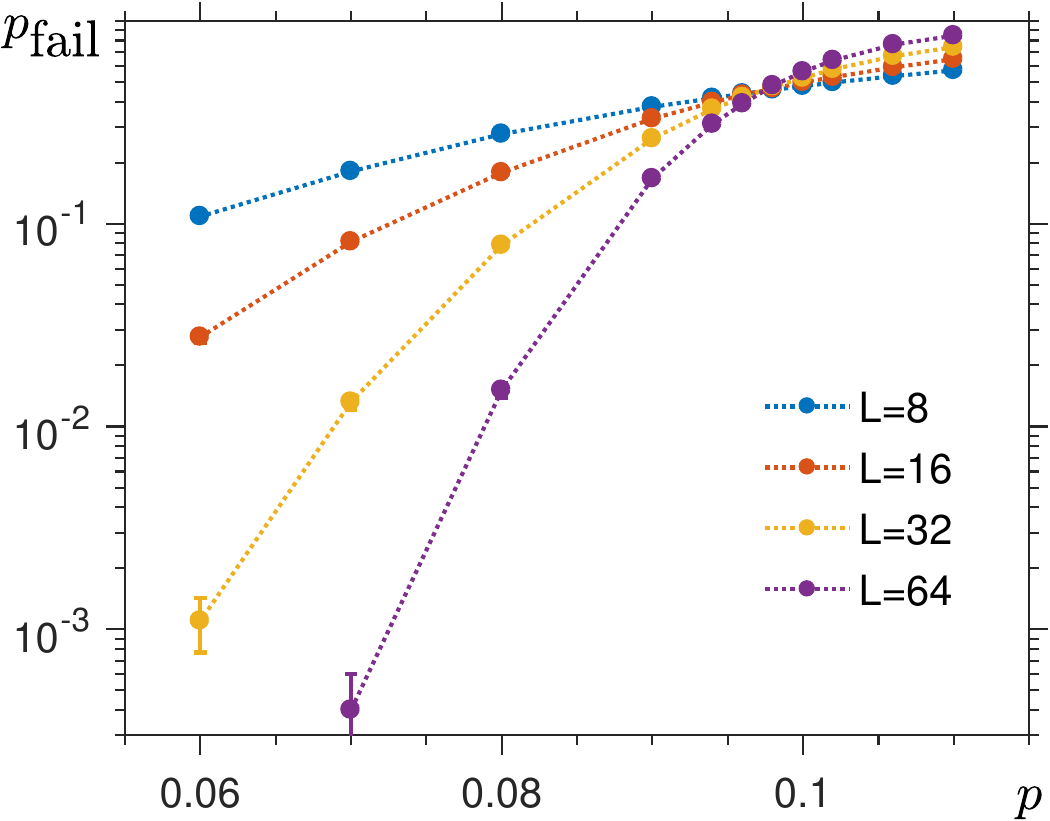}
\hspace*{12mm}
(b)\includegraphics[width=0.4\textwidth,height = .22\textheight]{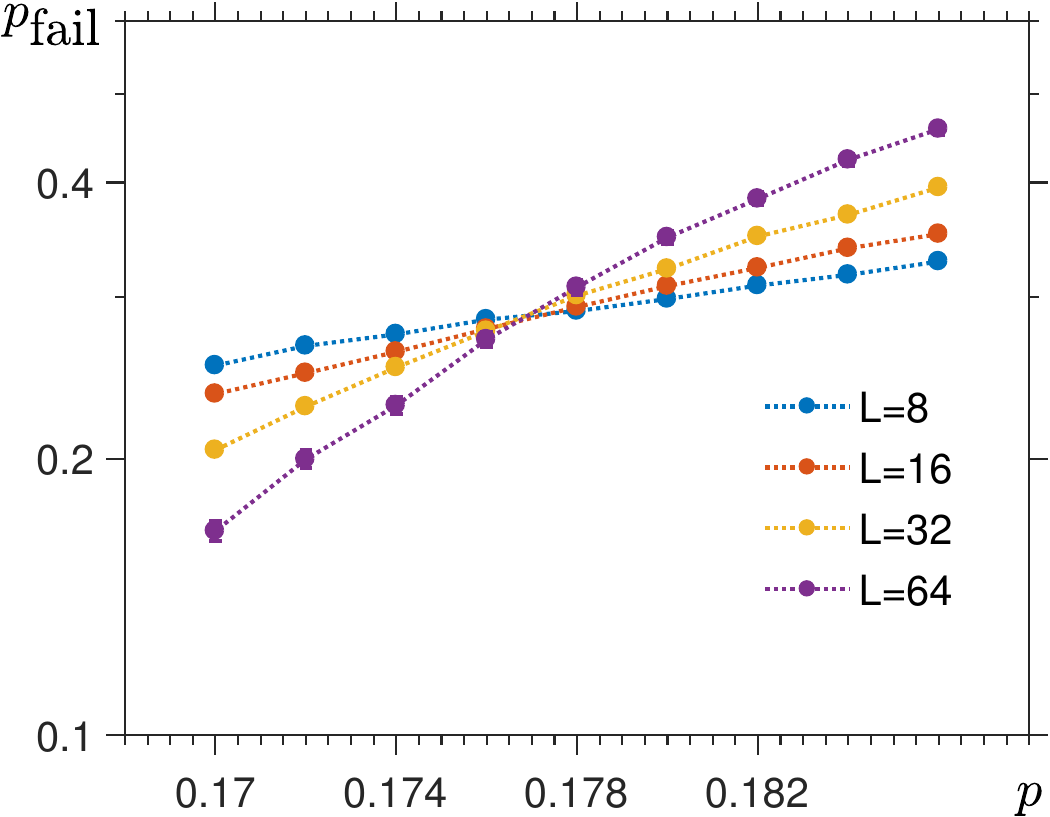}
\caption{
The decoding failure probability $p_{\textrm{fail}}(p,L)$ as a function of the noise strength $p$ and linear size $L$ of the lattice for:
(a) the Restriction Decoder (which uses the UF decoder) for the 2D color code on (a lattice dual to) the square-octagon lattice and
(b) the UF decoder for the 2D toric code on the $2$-square lattice.
}
\label{fig_uf}
\end{figure}

As we already mentioned, we can use any toric code decoder as a subroutine in the Restriction Decoder.
In particular, for the 0D point-like excitations we could use the UF decoder~\cite{Delfosse2017}.
Then, the runtime of the resulting Restriction Decoder would be almost linear in the number of qubits $N$, i.e., a worst case complexity of decoding would be $\mathcal O(N\alpha(N))$, where $\alpha(N)$ is a very slowly growing function of $N$ and for all practical purposes $\alpha(N)\leq 3$; see Ref.~\cite{Delfosse2017} for the details.

For an illustration, let us consider the 2D color code on (a lattice dual to) the square-octagon lattice $\mathcal{L}$ and the phase-flip noise, assuming perfect syndrome measurements; see Fig.~\ref{fig_sqoct}(a).
We choose the restricted lattices $\mathcal{L}_{RG}$ and $\mathcal{L}_{RB}$ in the same way as in Sec.~\ref{sec_sqoct}, i.e., we set $c^* = R$ and $\mathcal{C} = \{ G, B\}$.
We numerically estimate the threshold of the Restriction Decoder (which uses the UF decoder as a subroutine) to be $p'_{\textrm{2D}} \approx \psqoctUF$; see Fig.~\ref{fig_uf}(a).

We remark that we can estimate the color code threshold from the UF decoder threshold for the 2D toric code.
First, note that $\mathcal{L}_{RG}$ and $\mathcal{L}_{RB}$ are the same and we refer to them as the $2$-square lattices.
Since there is an error on the edge of the restricted lattice iff there is exactly one error on two triangles containing that edge in the original lattice $\mathcal{L}$, thus the effective noise strength for the toric code on the $2$-square lattice is
\begin{eqnarray}
f_{2\textrm{sq}}(p) &=& 2p(1-p).
\end{eqnarray}
We numerically find the thresholds of the UF decoder for the 2D toric code on the $2$-square lattice to be $q^{\textrm{0D}}_{2\textrm{sq}} \approx 17.7\%$; see Fig.~\ref{fig_uf}(b).
Finally, by using Eq.~(\ref{eq_threshold_estimate}) we estimate the threshold of the Restriction Decoder (which uses the UF decoder as a subroutine) to be $p'_{\textrm{2D}} \approx f^{-1}_{2\textrm{sq}}(q^{\textrm{0D}}_{2\textrm{sq}} )\approx \psqoctUF$, which is in agreement with our previous numerical estimation.

\subsection{Decoding the 2D color code without restricted lattices}
\label{sec_simplified_2D}

We can further simplify the Restriction Decoder for the 2D color code described in Sec.~\ref{sec_decoder_2D}.
Namely, we do not need to construct restricted lattices at all!
We illustrate this practical simplification in Fig.~\ref{fig_simple_nico}.

The first step of a simplified version of the Restriction Decoder is to find two subsets of edges $\widetilde{\rho_{RG}}, \widetilde{\rho_{RB}} \subseteq \face 1 {\mathcal{L}}$, whose $0$-boundaries respectively match the restricted syndromes $\sigma_{RG},\sigma_{RB}\subseteq \face 0 {\mathcal{L}}$, i.e., $\partial_1 \widetilde{\rho_{RG}} = \sigma_{RG}$ and $\partial_1 \widetilde{\rho_{RB}} = \sigma_{RB}$.
We can find $\widetilde{\rho_{RG}}$ and $\widetilde{\rho_{RB}}$ by using any decoder for the 2D toric code on the lattice $\mathcal{L}$ and the syndromes $\sigma_{RG}$ and $\sigma_{RB}$.
Note that this time $\widetilde{\rho_{RG}}$ and $\widetilde{\rho_{RB}}$ may contain vertices of color $B$ or $G$, respectively, since they are not limited to be only supported within the restricted lattices $\mathcal{L}_{RG}$ and $\mathcal{L}_{RB}$.

\begin{figure}[ht!]
\centering
(a)\includegraphics[height = 0.222 \textheight]{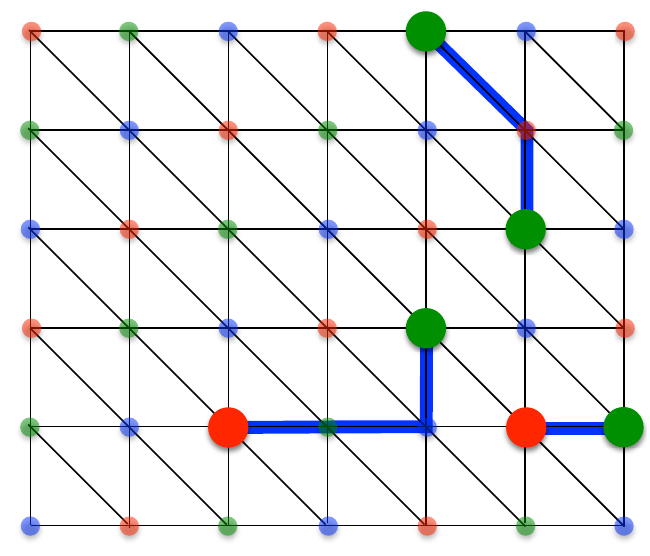}
\hspace*{12mm}
(b)\includegraphics[height = 0.22 \textheight]{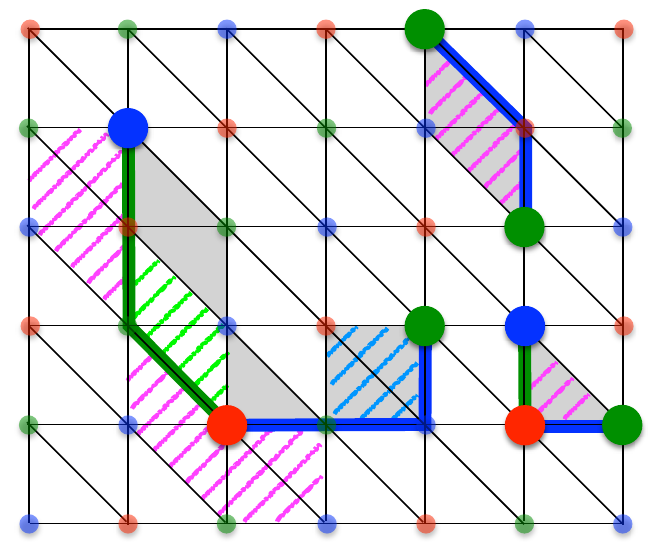}\\
\caption{
A simplified version of the Restriction Decoder for the 2D color code, which does not require restricted lattices.
We assume the same initial error $\epsilon\subseteq\face 2{\mathcal{L}}$ as in Fig.~\ref{fig_colorcode_2D}(b).
(a) We find a subset of edges $\widetilde{\rho_{RG}} \subseteq \face 1 {\mathcal{L}}$ (blue), whose $0$-boundary matches the restricted syndrome $\sigma_{RG}\subseteq \face 0 {\mathcal{L}}$, i.e., $\partial_1 \widetilde{\rho_{RG}} = \sigma_{RG}$.
Note that there is a $B$ vertex which belongs to $\widetilde{\rho_{RG}}$.
(b) The simplified decoder finds the color code correction $\phi \subseteq \face 2 {\mathcal{L}}$ by using a local lifting procedure for all the $R$ vertices of $\widetilde{\rho_{RG}} + \widetilde{\rho_{RB}}$, as well as for all the $B$ and $G$ vertices of $\widetilde{\rho_{RG}}$ and $\widetilde{\rho_{RB}}$, respectively.
We depict the output of the local lifting procedure for the $R$, $G$ and $B$ vertices as the faces hatched in magenta, light green and light blue, respectively.
Note that the initial error $\epsilon$ (shaded in grey) combined with the correction $\phi$ forms a stabilizer.
}
\label{fig_simple_nico}
\end{figure}

The second step of the simplified decoder is to use a local lifting procedure to find a correction $\phi\subseteq\face 2 {\mathcal{L}}$ from $\widetilde{\rho_{RG}} + \widetilde{\rho_{RB}}$.
This time, however, we apply the lifting procedure not only to all $R$ vertices of $\widetilde{\rho_{RG}} + \widetilde{\rho_{RB}}$, but also 
to all the $B$ and $G$ vertices of $\widetilde{\rho_{RG}}$ and $\widetilde{\rho_{RB}}$, respectively.
Similarly as before, for every vertex $v\in\facex R 0 {\widetilde{\rho_{RG}} + \widetilde{\rho_{RB}}}$ we find a subset of faces $\tau_v\subseteq \star 2 v$, such that its $1$-boundary locally matches $\widetilde{\rho_{RG}} + \widetilde{\rho_{RB}}$, i.e., $(\partial_2 \tau_v)\rest v = (\widetilde{\rho_{RG}} + \widetilde{\rho_{RB}})\rest v$.
Moreover, for every vertex $u\in\facex B 0 {\widetilde{\rho_{RG}}}$ we find $\tau_u\subseteq \star 2 u$ satisfying $(\partial_2 \tau_u)\rest u = \widetilde{\rho_{RG}}\rest u$, whereas for every vertex $w\in\facex G 0 {\widetilde{\rho_{RB}}}$ we find $\tau_w\subseteq \star 2 w$ satisfying $(\partial_2 \tau_w)\rest w = \widetilde{\rho_{RB}}\rest w$.

Finally, the output of the simplified decoder is
\begin{equation}
\phi = \sum_{v \in\facex R 0 {\widetilde{\rho_{RG}} + \widetilde{\rho_{RB}}}} \tau_v
+ \sum_{u \in\facex B 0 {\widetilde{\rho_{RG}}}} \tau_u+ \sum_{w \in\facex B 0 {\widetilde{\rho_{RG}}}} \tau_w.
\end{equation}
We leave to the readers to show that $\phi$ is a valid color code correction.

\subsection{Cellular automaton decoder of the 3D color code}
\label{sec_CA_decoder}

We note that it seems impossible to have a strictly local (in the sense of space and time) decoder for two-dimensional topological stabilizer codes, since their syndrome is identified with zero-dimensional objects and they always have one-dimensional string-like logical operators \cite{Bombin2014, Yoshida2011, Haah2013}.
For instance, in the 2D toric code one could have a pair of violated stabilizers separated by an arbitrarily large distance due to a string-like error.
If the decoder used only local information about the syndrome, it would not know the relative positioning of the pair of violated stabilizers and thus could not find a recovery operator.
This observation is consistent with all known decoding strategies, which rely on non-local information, either explicitly, as in the case of the MWPM algorithm~\cite{Dennis2002}, or implicitly, as in the case of the renormalization-group decoders~\cite{Harrington2004,Duclos-Cianci2010,Bravyi2013a,Anwar2013}.

It is possible, however, to construct a cellular automaton decoder of the color code in three dimensions, since one type of the syndrome forms 1D loop-like objects.
We emphasize that if one wanted a fully local decoder for both $X$ and $Z$ errors, then one would need to consider the 4D color code of type $k=2$.
For simplicity, we focus on decoding $Z$ errors in the 3D color code of type $k=2$, as it captures all the important aspects of cellular-automaton decoding.

Similarly as for the 2D case in Sec.~\ref{sec_simplified_2D}, we consider a simplified version of the Restriction Decoder, which does not use restricted lattices.
Let $\epsilon \subseteq \face 3 {\mathcal{L}}$ be the set of qubits affected by $Z$ errors and $\sigma = \bnd 3 1 \epsilon$  be the corresponding $X$-type syndrome.
First, we consider three restricted syndromes $\sigma_{RGB}, \sigma_{RGY}, \sigma_{RBY}\subseteq \face 1 {\mathcal{L}}$ and for each of them we use the Sweep Decoder from Ref.~\cite{Kubica2018toom} in the original lattice $\mathcal{L}$ to find $\widetilde{\rho_{RGB}},\widetilde{\rho_{RGY}},\widetilde{\rho_{RBY}}\subseteq \face 2 {\mathcal{L}}$.
Then, in order to find a correction $\phi \subseteq \face 3 {\mathcal{L}}$ we use a local lifting procedure.
As in the 2D case, $\widetilde{\rho_{RGB}}$ may contain vertices of color $Y$ since we do not use the restricted lattice $\mathcal{L}_{RGB}$ but the original lattice $\mathcal{L}$; similarly for $\widetilde{\rho_{RGY}}$ and $\widetilde{\rho_{RBY}}$.
Thus, we need to apply the local lifting procedure to all the $R$ vertices of $\widetilde{\rho_{RGB}} + \widetilde{\rho_{RGY}} + \widetilde{\rho_{RBY}}$, as well as to all the $Y$, $B$ and $G$ of $\widetilde{\rho_{RGB}}$, $\widetilde{\rho_{RGY}}$ and $\widetilde{\rho_{RBY}}$, respectively.

For the convenience of the reader we summarize a simplified version of the Restriction Decoder for the $Z$ errors in the 3D color code of type $k=2$ on the lattice $\mathcal{L}$.
We emphasize that this simplified decoder can be implemented via a cellular automaton, and thus is fully local.
\begin{itemize}
\item For every triple of colors $C \in \{ RGB, RGY, RBY \}$ use the Sweep Decoder on the lattice $\mathcal{L}$ for the restricted syndrome $\sigma_{C}$ in order to find $\widetilde{\rho_{C}}\subseteq\face 2 {\mathcal{L}}$.
\item Use the local lifting procedure for every $R$ vertex of $\widetilde{\rho_{RGB}} + \widetilde{\rho_{RGY}} + \widetilde{\rho_{RBY}}$, as well as for every vertex of color $RGBY\setminus C$ belonging to $\widetilde{\rho_{C}}$.
\end{itemize}
Since this simplified version of the Restriction Decoder only uses the initial lattice $\mathcal{L}$ (instead of three restricted lattices $\mathcal{L}_{RGB}$, $\mathcal{L}_{RGY}$ and $\mathcal{L}_{RBY}$), it is relatively easy to implement.

\subsection{Numerical estimates of the 3D color code threshold from the 3D toric code threshold}

\begin{figure}
\centering
(a)\includegraphics[width = .27\textwidth]{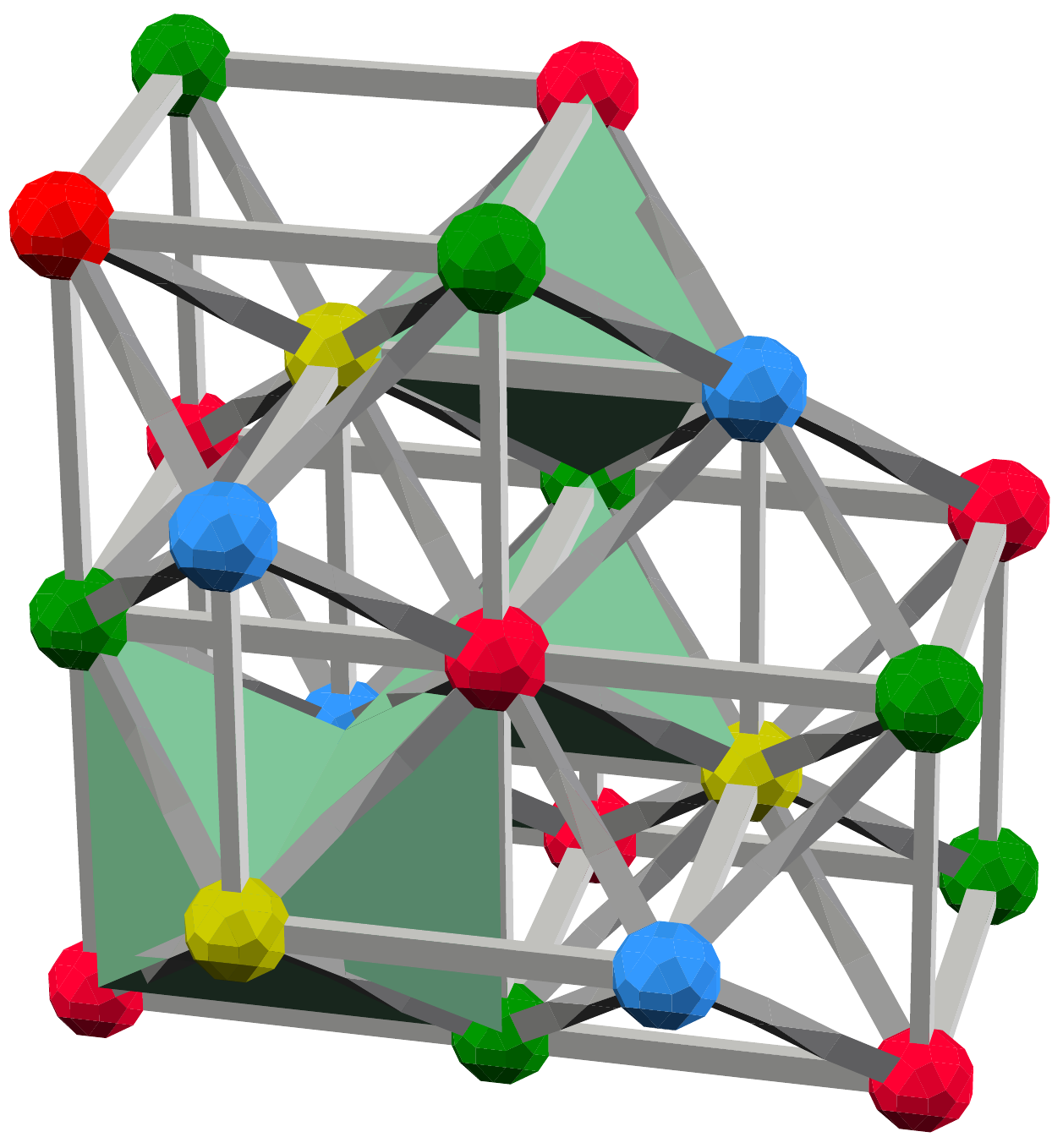}\hspace*{35mm}
(b)\includegraphics[width = .27\textwidth]{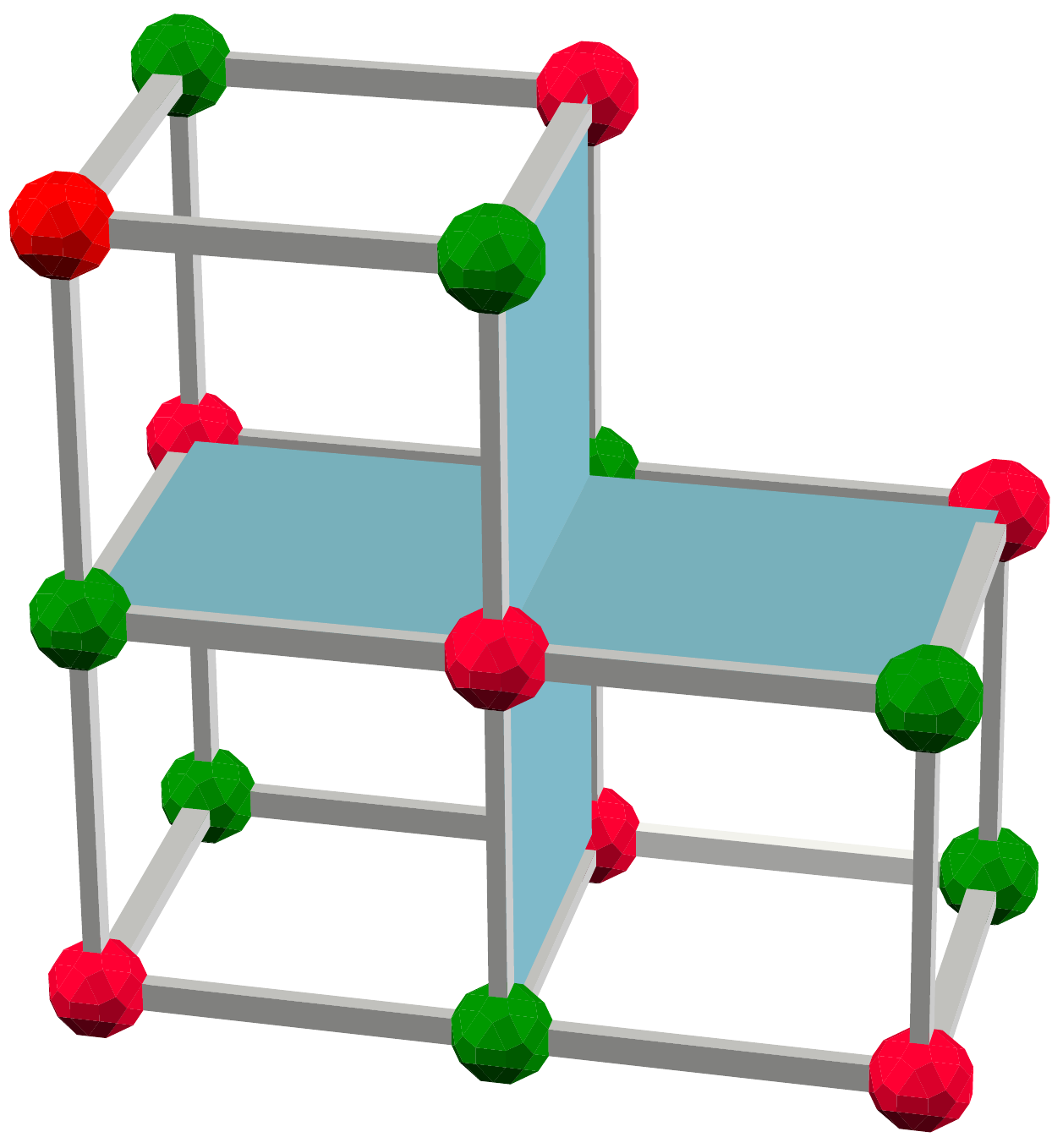}\\
(c)\includegraphics[width = .4\textwidth]{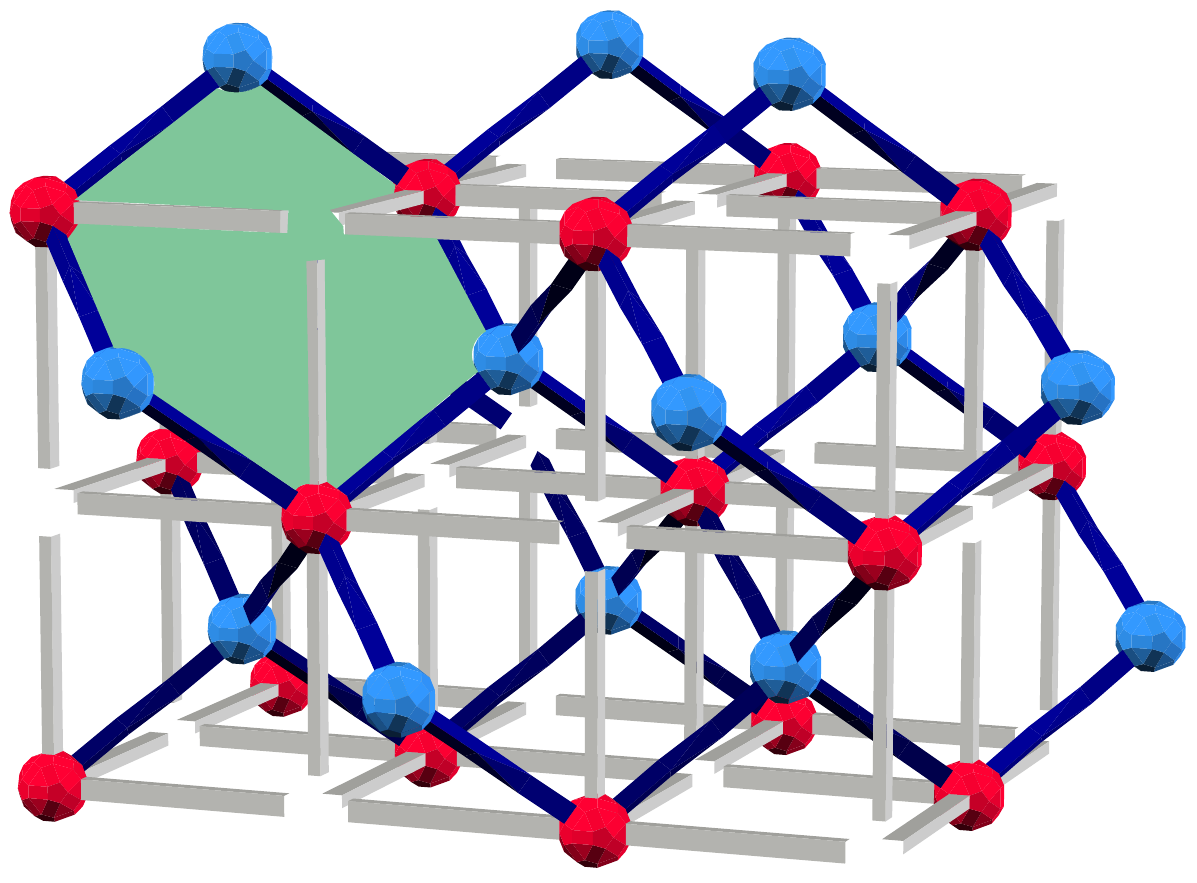}\hspace*{10mm}
(d)\includegraphics[width = .4\textwidth]{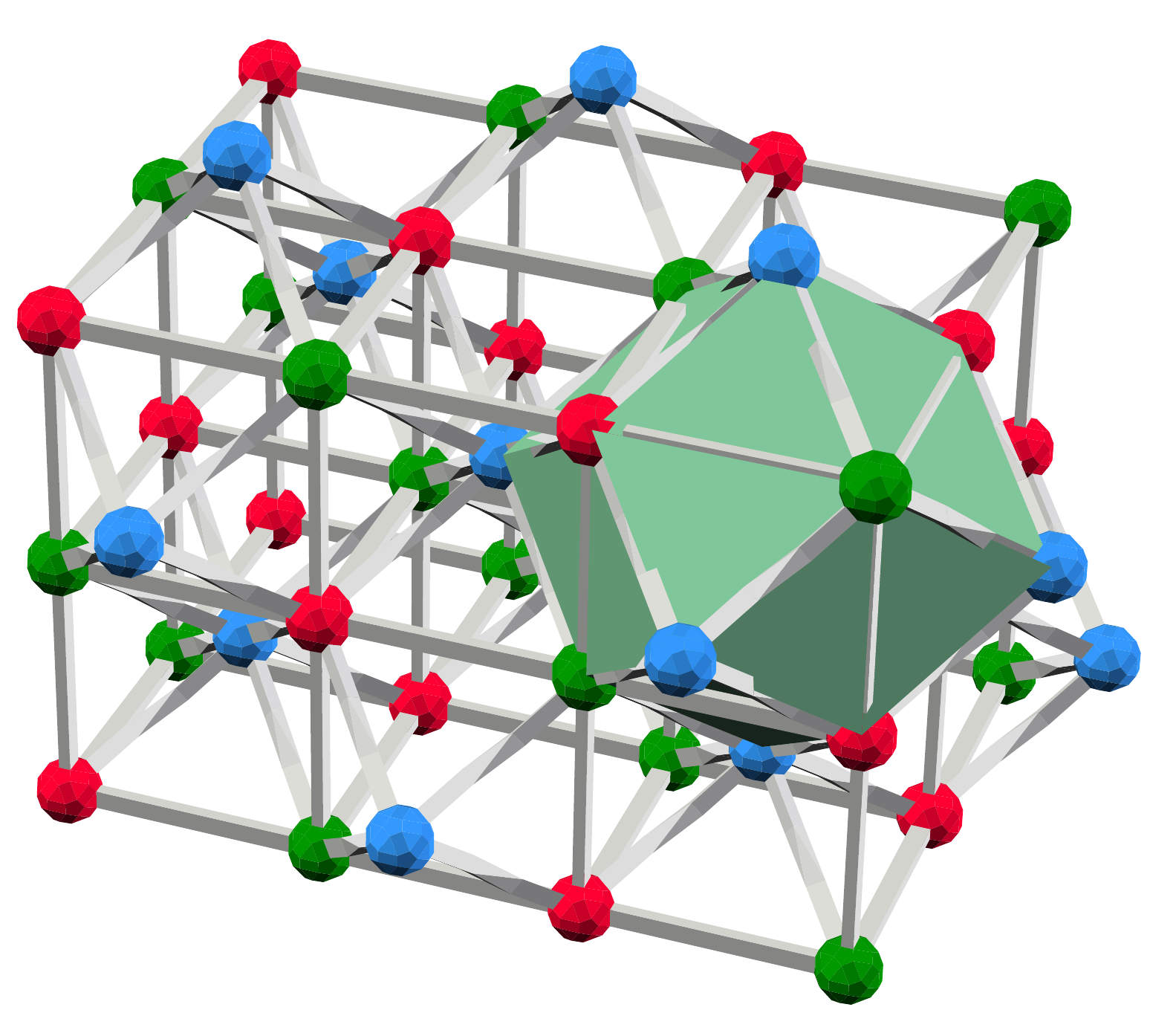}
\caption{
(a) The bcc lattice $\mathcal{L}$ is a lattice constructed from two interleaved cubic lattices (one corresponding to $R$ and $G$  vertices, and the other to $B$ and $Y$ vertices) by filling in tetrahedra (shaded in green).
The vertices of the bcc lattice are $4$-colorable.
(b) The restricted lattice $\mathcal{L}_{RG}$ forms the 3D cubic lattice.
$\mathcal{L}_{RG}$ is obtained from $\mathcal{L}$ by removing all $B$ and $Y$ vertices, as well as all the edges, faces and volumes containing them.
For each removed edge of color $BY$ we attach a square face (shaded in blue) to $\mathcal{L}_{RG}$.
(c) The restricted lattice $\mathcal{L}_{RB}$ forms the diamond cubic lattice, where every vertex is $4$-valent.
To make the figure readable, we keep one of the underlying cubic lattices (gray) and color the edges of $\mathcal{L}_{RB}$ in dark blue.
We shaded in green one of the hexagonal faces of $\mathcal{L}_{RB}$.
(d) The restricted lattice $\mathcal{L}_{RGB}$ is obtained from $\mathcal{L}$ by removing all $Y$ vertices, as well as all the edges, faces and volumes containing them.
For each removed vertex of color $Y$ we attach a volume (shaded in green) to $\mathcal{L}_{RGB}$.
We refer to $\mathcal{L}_{RGB}$ as the $\frac{3}{4}\textrm{bcc}$ lattice.
The figures were created using vZome available at {\texttt{http://vzome.com}}.}
\label{fig_restricted_lattice}
\end{figure}

\begin{figure}[ht!]
\centering
(a)\includegraphics[width=0.4\textwidth,height = .22\textheight]{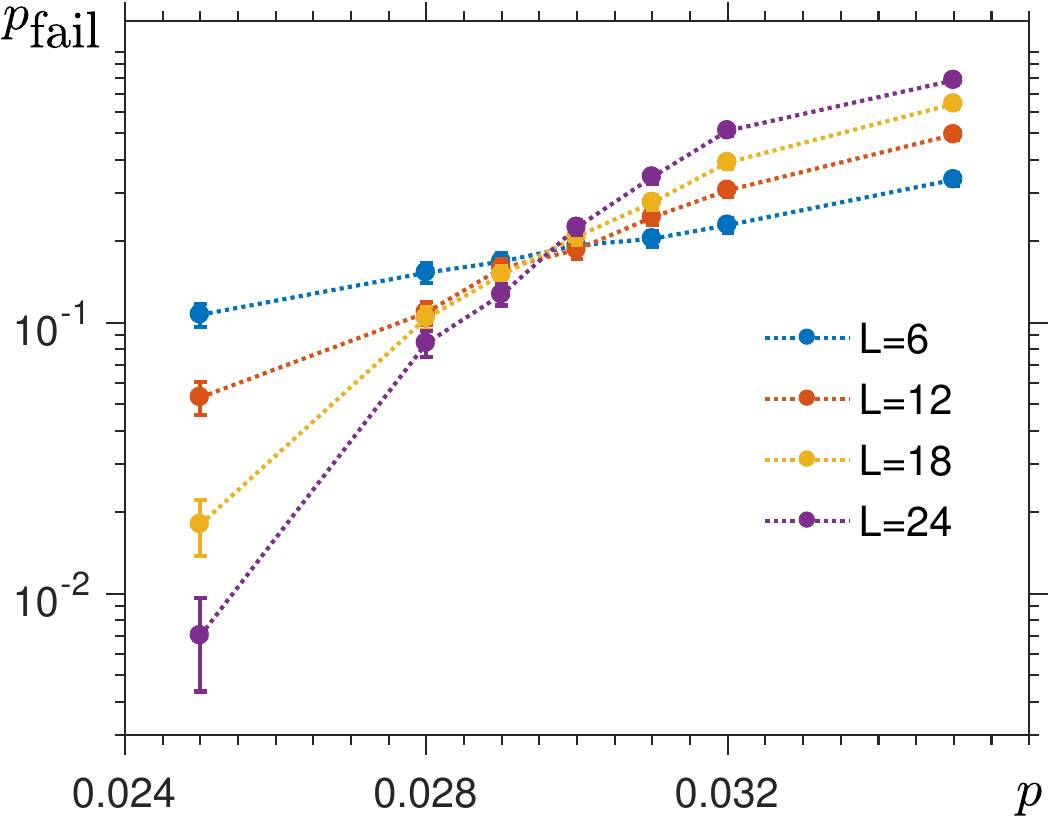}
\hspace*{12mm}
(b)\includegraphics[width=0.4\textwidth,height = .22\textheight]{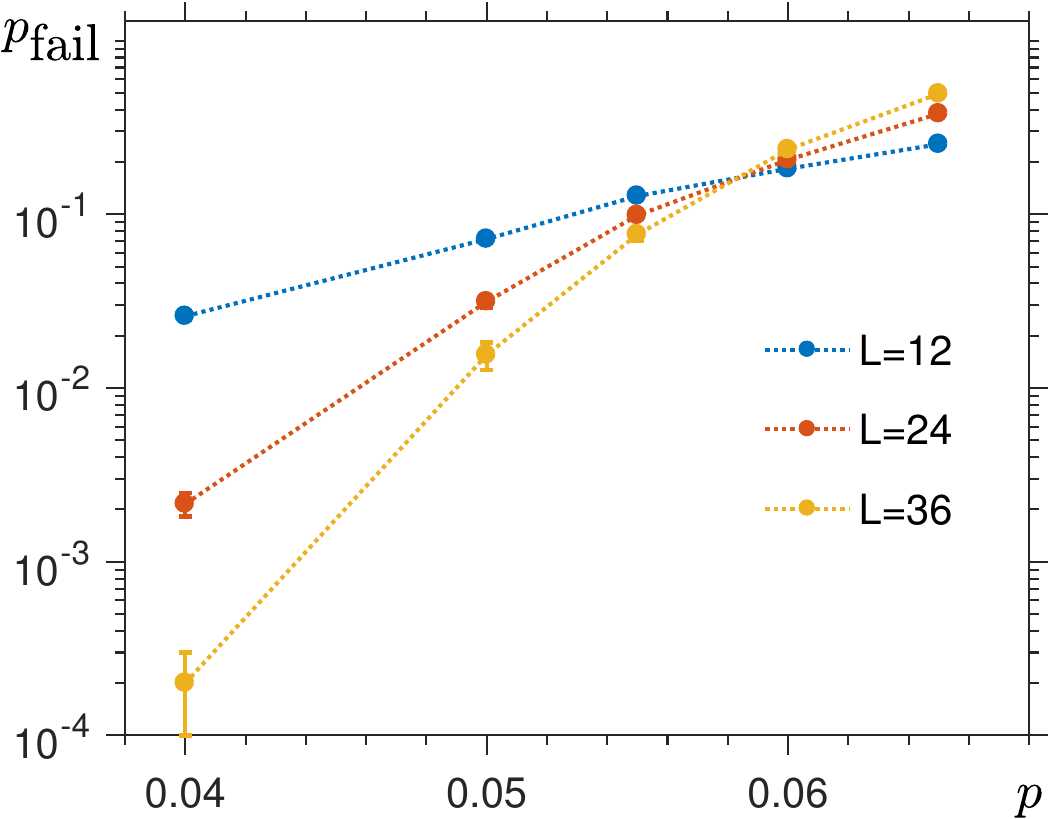}\\
\vspace*{4mm}
(c)\includegraphics[width=0.4\textwidth,height = .22\textheight]{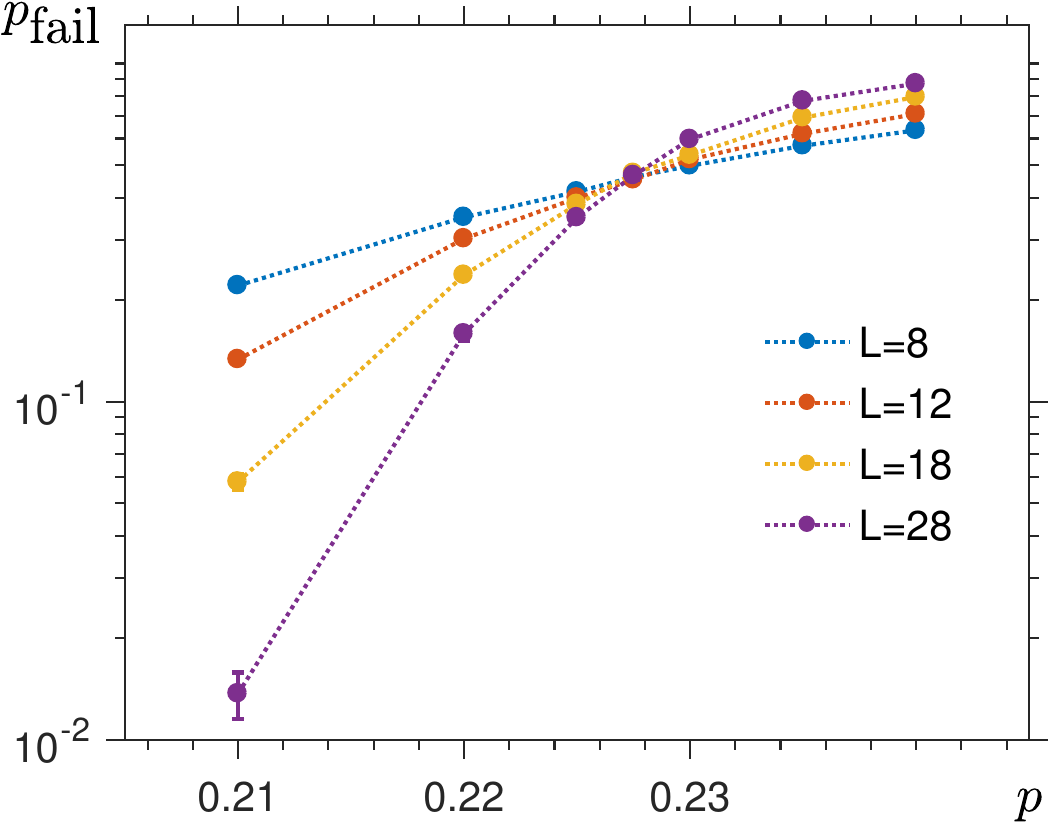}
\hspace*{12mm}
(d)\includegraphics[width=0.4\textwidth,height = .22\textheight]{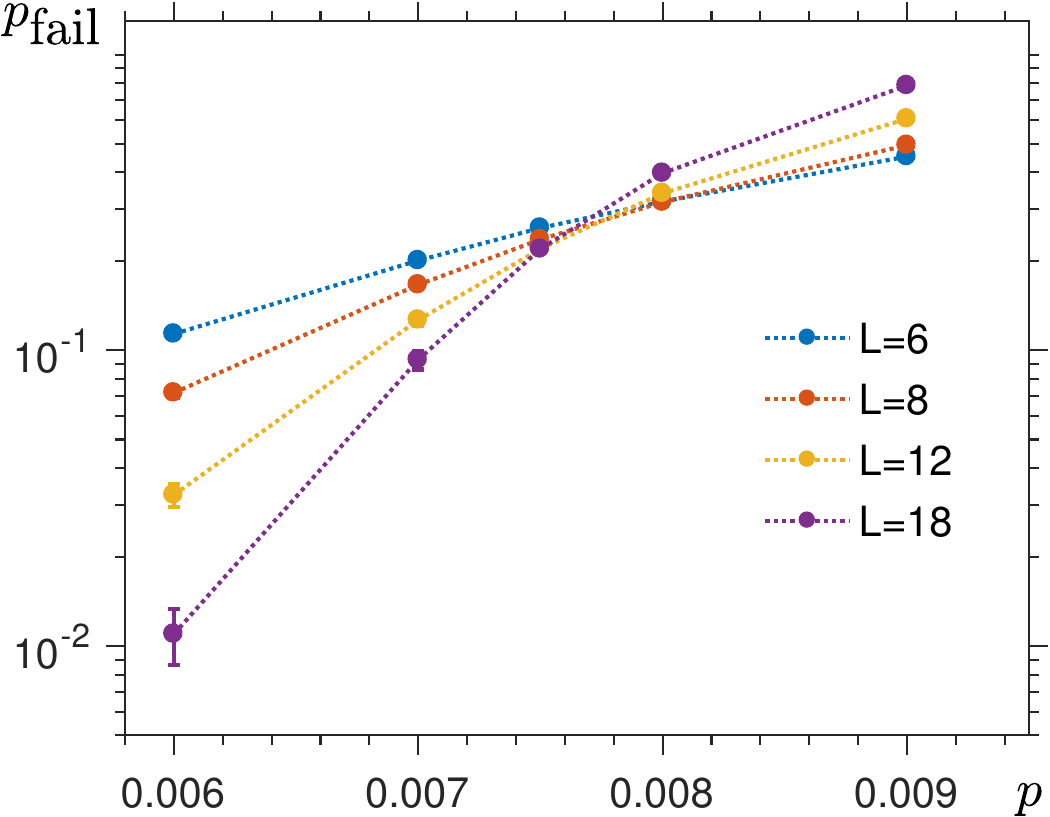}
\caption{
The decoding failure probability $p_{\textrm{fail}}(p,L)$ as a function of the noise strength $p$ and the linear size $L$ of the lattice for: (a)-(c) the 3D toric code and (d) the 3D color code.
In (a) and (b) we use the MWPM algorithm for 0D point-like excitations on the cubic and the diamond cubic lattices.
In (c) we use the Sweep Decoder for 1D loop-like excitations on the $\frac{3}{4}\textrm{bcc}$ lattice.
In (d) we use the Restriction Decoder for the 0D point-like excitations of the 3D color code on the bcc lattice.
}
\label{fig_thresholds}
\end{figure}

We now consider the 3D color code of type $k=1$ on the bcc lattice $\mathcal{L}$; see Fig.~\ref{fig_restricted_lattice}(a).
We assume the phase-flip noise and perfect syndrome extraction, and set $c^* = R$ and $\mathcal{C} = \{ G, B, Y\}$.
Recall that qubits are placed on tetrahedral volumes of $\mathcal{L}$ and the $X$-type syndrome can be viewed as 0D point-like excitations.
Note that the restricted lattice $\mathcal{L}_{RG}$ forms the cubic lattice, whereas the restricted lattices $\mathcal{L}_{RB}$ and $\mathcal{L}_{RY}$ form the cubic diamond lattices; see Fig.~\ref{fig_restricted_lattice}(b)(c).
We observe that for any $C\in\mathcal C$ there is an error on the edge of the restricted lattice $\mathcal{L}_{C^*}$, where $C^* = C\sqcup\{c^*\}$, iff there is an odd number of errors on tetrahedra containing that edge in the original lattice $\mathcal{L}$.
Thus, we find the effective noise strength for the toric code on the cubic and the diamond cubic lattices to be
\begin{eqnarray}
f_{\textrm{cub}}(p) &=& 4p(1-p)^3+ 4p^3(1-p),\\
f_{\textrm{dia}}(p) &=& 6p(1-p)^5+ 20p^3(1-p)^3 + 6p^5(1-p).
\end{eqnarray}
We numerically find the thresholds of the MWPM algorithm for the 3D toric code on the cubic and the diamond cubic lattices to be $q^{\textrm{0D}}_{\textrm{cub}} \approx 2.95\%$ and $q^{\textrm{0D}}_{\textrm{dia}} \approx 5.8\%$, respectively; see Fig.~\ref{fig_thresholds}(a)(b).
Using Eq.~(\ref{eq_threshold_estimate}), we arrive at an estimate of the Restriction Decoder threshold for the 3D color code on the bcc lattice and 0D point like excitations
\begin{equation}
\label{eq_cc_threshold_0D}
p^{\textrm{0D}}_{\textrm{bcc}} \approx \min \left\{ f_{\textrm{cub}}^{-1}(q^{\textrm{0D}}_{\textrm{cub}}), 
f_{\textrm{dia}}^{-1}(q^{\textrm{0D}}_{\textrm{dia}})\right\} \approx 0.75\%.
\end{equation}
We verify tightness of the estimate of the Restriction Decoder threshold by directly implementing the Restriction Decoder based on the MWPM algorithm and the local lifting procedure $\mathtt{Lift}$, as described in Sec.~\ref{sec_restriction_decoder}.
We find the threshold to be $p^{\textrm{0D}}_{\textrm{bcc}}\approx 0.77\%$, which is in agreement with Eq.~(\ref{eq_cc_threshold_0D}); see Fig.~\ref{fig_thresholds}(d).
This value compares favorably with the clustering decoder threshold $p^{\textrm{0D}}_{\textrm{clu}}\approx 0.46\%$ \cite{Brown2015} (which was estimated for the 3D color code but not on the bcc lattice).

We can also estimate the Restriction Decoder threshold for the bit-flip noise.
In that case the $Z$-type syndrome forms 1D loop-like excitations, as it is some subset of edges of $\mathcal{L}$.
We set $c^* = R$ and $\mathcal{C} = \{ GB, GY, BY\}$.
Observe that all the restricted lattices $\mathcal{L}_{RGB}$, $\mathcal{L}_{RGY}$ and $\mathcal{L}_{RBY}$ form the same lattice, which we call the $\frac{3}{4}\textrm{bcc}$ lattice; see Fig.~\ref{fig_restricted_lattice}(d).
Moreover, for any $C\in\mathcal C$ there is an error on the face of the restricted lattice $\mathcal{L}_{C^*}$, where $C^* = C\sqcup\{c^*\}$, iff there is exactly one error on two neighboring tetrahedra containing that face in the original lattice $\mathcal{L}$.
Thus, the effective noise strength for the toric code on the $\frac{3}{4}\textrm{bcc}$ lattice is 
\begin{eqnarray}
f_{\frac{3}{4}\textrm{bcc}}(p) &=& 2p(1-p).
\end{eqnarray}
We numerically find the threshold of the Sweep Decoder from Ref.~\cite{Kubica2018toom} for the toric code on the $\frac{3}{4}\textrm{bcc}$ to be
$q^{\textrm{1D}}_{\frac{3}{4}\textrm{bcc}} \approx 22.75\%$; see Fig.~\ref{fig_thresholds}(c).
Using Eq.~(\ref{eq_threshold_estimate}), we arrive at an estimate of the Restriction Decoder threshold for the 3D color code on the bcc lattice and 1D loop-like excitations
\begin{equation}
p^{\textrm{1D}}_{\textrm{bcc}} \approx \frac{1}{2}\left(1 - \sqrt{1-2 q^{\textrm{1D}}_{\frac{3}{4}\textrm{bcc}}}\right)
\approx 13.1\%.
\end{equation}

\section{Discussion}

In our work we provide an efficient solution to an open problem of decoding the color code in $d\geq 2$ dimensions.
Namely, we introduce the Restriction Decoder, which combines any $d$-dimensional toric code decoder with a local lifting procedure to find a recovery operation for the color code.
Furthermore, we prove that the Restriction Decoder successfully corrects errors in the color code if and only if the corresponding toric code decoding succeeds.
As an immediate consequence, we obtain that the performance of the Restriction Decoder is determined by the performance of the toric code decoder that we use, minimized over the toric codes on the restricted lattices that we select.
We remark that the Restriction Decoder for the color code in $d\geq 3$ dimensions can be implemented via a cellular automaton if one uses as the toric code decoder the Sweep Decoder introduced in the accompanying article~\cite{Kubica2018toom}.

We numerically estimate the Restriction Decoder threshold for the color code in two and three dimensions against the phase-flip and bit-flip noise, assuming perfect syndrome extraction.
The Restriction Decoder threshold $p_{\textrm{2D}} \approx \psqoct$ for the 2D color code on (a lattice dual to) the square-octagon lattice is on a par with the thresholds $9.9\sim 10.3\%$ of the leading efficient decoders for the 2D toric code on the square lattice~\cite{Dennis2002,Delfosse2017}, as well as with the optimal decoding threshold $10.9\%$ for the 2D color code~\cite{Katzgraber2009}.
We remark that the Restriction Decoder surpasses the highest previously reported 2D color code threshold, which was achieved by heuristic neural-network decoding~\cite{Maskara2018}. 
We also estimate the Restriction Decoder thresholds for correcting 0D point-like and 1D loop-like excitations in the 3D color code on the bcc lattice to be $p^{\textrm{0D}}_{\textrm{bcc}} \approx 0.77\%$ and $p^{\textrm{1D}}_{\textrm{bcc}} \approx 13.1\%$.
Note that those values should be compared with the optimal decoding thresholds $p^{\textrm{0D}}_{\textrm{opt}} \approx 1.9\%$ and $p^{\textrm{1D}}_{\textrm{opt}} \approx 27.6\%$ estimated via statistical-mechanical mappings~\cite{Kubica2017}.

We emphasize that there is a lot of structure in the color code decoding problem, mainly due to the colorability of the lattice (see Ref.~\cite{Vasmer2022} for a local modification of the color code and its lattice).
This combinatorial structure makes the problem challenging but at the same time may allow for color code decoders, which do not rely on toric code decoding as a subroutine.
In particular, incorporating the colorability of the lattice into the Sweep Rule~\cite{Kubica2018toom} could lead to native cellular-automaton decoders for the color code with error-correction thresholds higher than the ones we numerically estimated.

Our discussion assumes perfect syndrome extraction, however in a realistic scenario one needs to account for measurement errors.
To decode 0D point-like excitations in the presence of measurement errors, one needs to perform multiple rounds of syndrome extraction and then use the collected data to find out the appropriate correction.
In order to decode 1D loop-like (or higher-dimensional) excitations we expect to use a version of the Restriction Decoder based on the cellular-automaton Greedy Sweep Decoder~\cite{Kubica2018toom,Vasmer2021}, which only requires local (possibly faulty) syndrome information.
We do not perceive any fundamental difficulties with incorporating a possibility of measurement errors into the Restriction Decoder, as well as incorporating the boundaries; see, for instance, the follow-up work~\cite{Chamberland2020}.

\subsection*{Acknowledgements}
We thank T. Jochym-O'Connor for his comments on the draft of our work.
A.K. acknowledges funding provided by the Simons Foundation through the ``It from Qubit'' Collaboration.
Research at Perimeter Institute is supported by the Government of Canada through Industry Canada and by the Province of Ontario through the Ministry of Research and Innovation.

\bibliography{cc_decoder_pruned}

\end{document}